\documentclass{aa}
\usepackage{graphicx}

\def\jh{\mbox{$(J-H)$}}

\def\mMJ{\mbox{$(m-M)_J$}}
\def\mMo{\mbox{$(m-M)_O$}}
\def\ebv{\mbox{$E(B-V)$}}
\def\ejh{\mbox{$E(J-H)$}}
\def\rc{\mbox{$R_{\rm core}$}}
\def\rl{\mbox{$R_{\rm lim}$}}
\def\rt{\mbox{$R_{\rm tidal}$}}
\def\ms{\mbox{$M_\odot$}}
\def\ds{\mbox{$d_\odot$}}
\def\dgc{\mbox{$d_{\rm GC}$}}
\def\jj{\mbox{$J$}}
\def\hh{\mbox{$H$}}
\def\ks{\mbox{$K_s$}}

\def\mobs{\mbox{$m_{\rm obs}$}}

\def\kms{\mbox{$\rm km\,s^{-1}$}}
\def\tr{\mbox{$t_{\rm rel}$}}
\def\tcr{\mbox{$t_{\rm cr}$}}

\def\avgE{\mbox{$\overline{\Delta\ebv}$}}

\begin{document}

\title{Methods for improving open cluster fundamental parameters applied to M\,52 and NGC\,3960}

\author{C. Bonatto\inst{1} \and E. Bica\inst{1}}

\offprints{Ch. Bonatto}

\institute{Universidade Federal do Rio Grande do Sul, Instituto de F\'\i sica, 
CP\,15051, Porto Alegre 91501-970, RS, Brazil\\
\email{charles@if.ufrgs.br, bica@if.ufrgs.br}
\mail{charles@if.ufrgs.br} }

\date{Received --; accepted --}

\abstract{}
{We derive accurate parameters related to the CMD, structure and dynamical state of M\,52 and 
NGC\,3960, whose fields are affected by differential reddening. Previous works estimated their 
ages in the ranges $35 - 135$\,Myr and $0.5 - 1.0$\,Gyr, respectively.}
{\jj, \hh\ and \ks\ 2MASS photometry with errors $<0.2$\,mag is used to build CMDs, radial density 
profiles, luminosity and mass functions, and correct for differential reddening. Field-star 
decontamination is applied to uncover the cluster's intrinsic CMD morphology, and colour-magnitude 
filters are used to isolate stars with high probability of being cluster members.}
{The differential-reddening corrected radial density profile of M\,52 follows King's law with core 
and limiting radii of $\rm\rc=0.91\pm0.14\,pc$ and $\rm\rl=8.0\pm1.0\,pc$. NGC\,3960 presents an 
excess of the stellar density over King's profile ($\rm\rc=0.62\pm0.11\,pc$ and $\rm\rl=6.0\pm0.8\,pc$) 
at the center. The tidal radii of M\,52 and NGC\,3960 are $\rt=13.1\pm2.2$\,pc and $\rt=10.7\pm3.7$\,pc.
Cluster ages of M\,52 and NGC\,3960 derived with Padova isochrones are constrained to $60\pm10$\,Myr 
and $1.1\pm0.1$\,Gyr. In M\,52 the core MF ($\chi_{\rm core}=0.89\pm0.12$) is flatter than the halo's 
($\chi_{\rm halo}=1.65\pm0.12$). In NGC\,3960 they are $\chi_{\rm core}=-0.74\pm0.35$ and 
$\chi_{\rm halo}=1.26\pm0.26$. The mass locked up in MS/evolved stars in M\,52 is $\sim1200$\,\ms, and 
the total 
mass (extrapolated to 0.08\,\ms) is $\sim3800$\,\ms. The total mass in NGC\,3960 is $\sim1300$\,\ms.}
{Compared to open clusters in different dynamical states studied with similar methods, the core and 
overall parameters of M\,52 are consistent with an open cluster more massive than 1\,000\,\ms\ and 
$\sim60$\,Myr old, with some mass segregation in the inner region. The core of NGC\,3960 is in an
advanced dynamical state with strong mass segregation in the core/halo region, while the somewhat flat  
overall MF ($\chi\approx1.07$) suggests low-mass star evaporation. The excess stellar density in the
core may suggest post-core collapse. The dynamical evolution of NGC\,3960 may have been accelerated by 
the tidal Galactic field, since it lies $\approx0.5$\,kpc inside the Solar circle.}


\titlerunning{Fundamental parameters of M\,52 and NGC\,3960}

\authorrunning{C. Bonatto and E. Bica}

\maketitle

\section{Introduction}
\label{intro}

Open clusters (OCs) can be used as test-beds of molecular cloud fragmentation, star formation, and stellar 
and dynamical evolution models. They are excellent probes of the Galactic disc structure (Janes \&
Phelps \cite{JP94}; Friel \cite{Friel95}; Bonatto et al. \cite{BKBS06}). This follows from the relative
simplicity in age and distance estimates, at least for bright clusters. Accurate OC parameters such as
core and limiting radii, age, mass, density, and relaxation times are essential to constrain theoretical 
models. However, because most of the OCs lie close to the disk, and thus are more affected both
by reddening and field-star contamination (Bonatto et al. \cite{BKBS06} and references therein), this kind 
of analysis becomes restricted to the more populous OCs and/or those located a few kpc from the Sun. 

According to the WEBDA\footnote{\em http://obswww.unige.ch/webda} OC database (Mermilliod \cite{Merm1996}) 
the current number of OCs with known parameters (such as coordinates, age, distance from the Sun, and
reddening) amounts to $\sim700$. The catalogue of Dias et al. (\cite{Dias2002}) includes 1756 optically 
visible OCs and candidates. More recently Bonatto et al. (\cite{BKBS06}) simulated observational completeness 
in different directions throughout the Galaxy and estimated that $\sim50$\% of Trumpler types I to III OCs 
remain undetected in the solar neighbourhood (distance from the Sun $\ds\leq1.3$\,kpc). A large fraction of the
faint and/or poorly populated OCs end up drowned in the stellar background, particularly in directions 
intercepting the bulge. 

\begin{figure*}
\begin{minipage}[b]{0.50\linewidth}
\includegraphics[width=\textwidth]{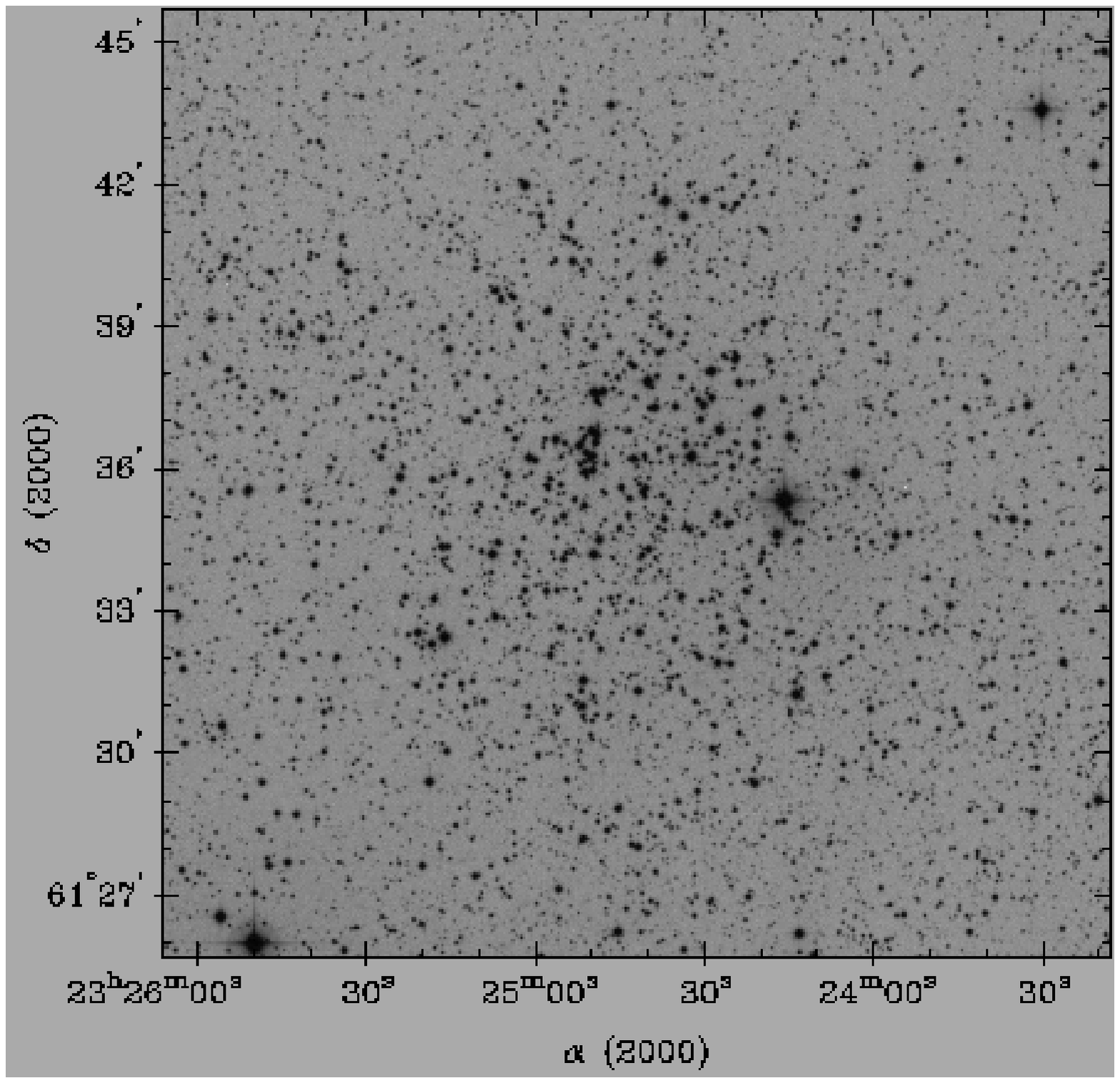}
\end{minipage}\hfill
\begin{minipage}[b]{0.50\linewidth}
\includegraphics[width=\textwidth]{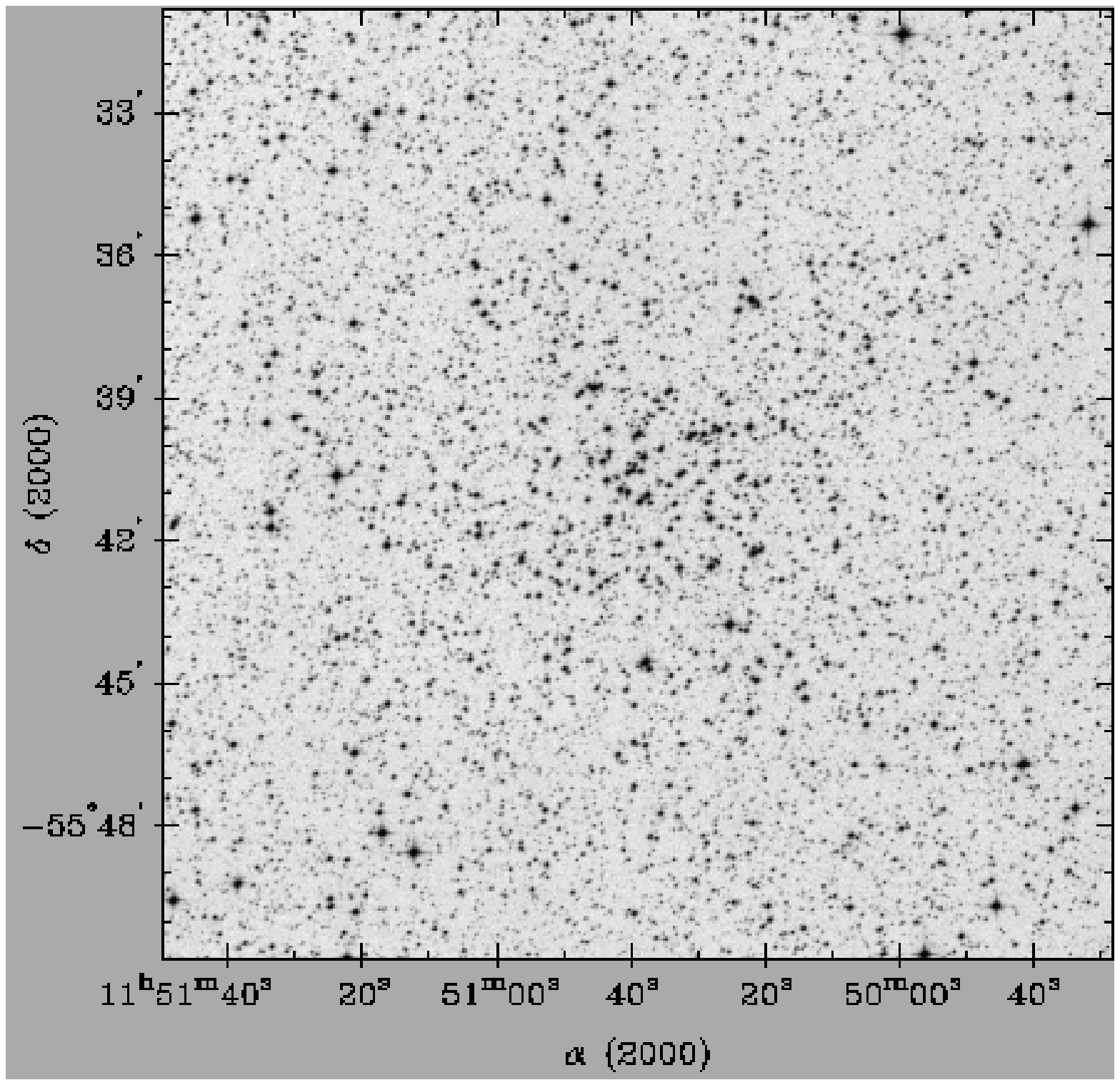}
\end{minipage}\hfill
\caption[]{$\rm20\arcmin\times20\arcmin$ XDSS R images of M\,52 (left panel) and NGC\,3960 
(right panel).}
\label{fig1}
\end{figure*}

Several decades of studies have provided fundamental parameters for a considerable fraction of the
catalogued OCs (e.g. Lyng\aa\ \cite{Lynga82}; WEBDA; Dias et al. \cite{Dias2002}; Kharchenko et al.
\cite{Kharchenko05}; Bonatto et al. \cite{BKBS06}). Some clusters have been studied twice or more
and a disagreement of a factor of $\sim2$ in parameters is not uncommon. Discrepancies arise mostly 
from different analytical methods dealing with optical or IR photometry, Colour-Magnitude Diagrams 
(CMDs), isochrone sets, photometric systems, etc. 

Recently we started a program to derive fundamental parameters of OCs with near-IR photometry
and a series of analytical procedures developed to minimise field-star contamination of CMDs. They 
have proven to be effective for parameter determination for a wide class of OCs.
For spatial and photometric uniformity we work with \jj, \hh\ and \ks\ 2MASS\footnote {The Two 
Micron All Sky Survey, All Sky data release (Skrutskie et al. \cite{2mass1997}), available at 
{\em http://www.ipac.caltech.edu/2mass/releases/allsky/}} photometry. The 2MASS Point Source 
Catalogue (PSC) is uniform reaching relatively faint magnitudes covering nearly all the sky, 
allowing a proper background definition for clusters with large angular sizes (e.g.
Bonatto, Bica \& Santos Jr. \cite{BBS2005}; Bonatto, Bica \& Pavani \cite{BBP2004}). 

The use of field-star decontamination and colour-magnitude filters have produced more robust 
parameters (e.g. Bonatto \& Bica \cite{BB2005}; Bonatto et al. \cite{BBOB06}). In particular, 
field-star decontamination constrains the age more, especially for low-latitude OCs (Bonatto 
et al. \cite{BKBS06}). These procedures have proven useful in the analysis also of faint and/or distant 
OCs (Bica, Bonatto \& Dutra \cite{BBD03}; Bonatto, Bica \& Dutra \cite{BBD04}; Bica \& Bonatto \cite{LowC05}).

Another potential source of observational uncertainties in OC parameters is the presence of differential 
reddening. Spatial variation of reddening may affect the age determination by isochrone fitting and,
consequently, the distance from the Sun. This in turn could affect the absolute values of core and
limiting radii, and cluster mass.

Our goals in this paper are twofold: {\em (i)} add to the previous techniques a procedure to take into
account the differential reddening, and {\em (ii)} derive more constrained parameters for two  
OCs of different ages. As targets we selected the relatively bright OCs M\,52 and NGC\,3960 that in the
literature have uncertain age determinations. The age of M\,52 has been estimated in the range $35 - 
135$\,Myr (Sect.~\ref{M52}), while that of NGC\,3960 varies from $0.5$ to $1.0$\,Gyr (Sect.~\ref{N3960}). 
Previous works have shown evidence of differential reddening in the fields of both clusters (e.g. Pandey 
et al. \cite{Pandey2001}; Prisinzano et al. \cite{Prisinzano04}). Besides reddening, age and distance from 
the Sun, in this paper we derive the core and limiting radii, and the stellar mass stored in
the core and halo subsystems. The latter parameters for NGC\,3960 are derived for the first time.
 
This paper is organized as follows. In Sect.~\ref{M52N3960} we review previous findings on M\,52 and NGC\,3960. 
In Sect.~\ref{2mass} we present the 2MASS data, subtract the field-star contamination, estimate the differential
reddening, derive fundamental cluster parameters, and analyse the cluster density structure. In Sect.~\ref{MF}
we derive luminosity and mass functions (LFs and MFs), and compute stellar content properties. In Sect.~\ref{dyna} 
we discuss the dynamical states of both OCs. In Sect.~\ref{CWODS} we use diagnostic diagrams to analyse the
structure and dynamical state of M\,52 and NGC\,3960 as compared to other OCs. Concluding remarks are 
given in Sect.~\ref{Conclu}. 

\section{The target open clusters M\,52 and NGC\,3960}
\label{M52N3960}

\subsection{M52}
\label{M52}

M\,52 (NGC\,7654) is a prominent OC located in the direction of Cassiopea, in an interarm 
region (Fenkart \& Schr\"oder \cite{FS85}). Its field is affected by a significant amount of variable 
reddening (e.g. Danford \& Thomas \cite{DT81}) and field star contamination (Sect.~\ref{FSD}), consistent
with its low galactic latitude (Table~\ref{tab1}). The supergiant (SG) star $BD+60^{\circ}2532$ of spectral 
type F7\,Ib with $V=8.22$ and $B-V=1.16$ that is projected $\approx3.1\arcmin$ from the cluster center 
(Sect.~\ref{2mass}) is a probable cluster member (Schmidt \cite{Schmidt84}). Other bright stars are projected in
the field of M\,52. These features appear in the $\rm20\arcmin\times20\arcmin$ DSS\footnote{Extracted 
from the Canadian Astronomy Data Centre (CADC), at \em http://cadcwww.dao.nrc.ca/} R image (left panel
of Fig.~\ref{fig1}). Below we summarise previous works on M\,52.

\begin{table*}
\caption[]{General data on M\,52 and NGC\,3960}
\label{tab1}
\renewcommand{\tabcolsep}{0.3mm}
\renewcommand{\arraystretch}{1.5}
\begin{tabular}{lccccccccccccccc}
\hline\hline
&\multicolumn{5}{c}{WEBDA}&\multicolumn{9}{c}{Present work}\\
\cline{2-6}\cline{8-16}
Cluster&$\alpha(2000)$&$\delta(2000)$&Age&\ebv&\ds&&$\alpha(2000)$&$\delta(2000)$&$\ell$&$b$&Age&$\ebv_O$&\ds&\dgc&\dgc \\
&(hms)&($^\circ\arcmin\arcsec$)&(Myr)&&(kpc)&&(hms)&($^\circ\arcmin\arcsec$)&($^\circ$)&($^\circ$)&(Myr)&&(kpc)&(kpc)&(kpc)\\
(1)&(2)&(3)&(4)&(5)&(6)&&(7)&(8)&(9)&(10)&(11)&(12)&(13)&(14)&(15)\\
\hline
M\,52&23:24:31&$+$61:35:29    &58&0.65&1.42&&23:24:42&$+$61:35:42&112.81&$+0.44$&$60\pm10$&$0.58^{0.93}_{0.50}$&1.4&8.7&7.9\\
NGC\,3960&11:50:33&$-$55:40:24&664&0.30&2.26&&11:50:38.9&$-$55:40:48&294.38&$+6.18$&$1100\pm100$&$0.13^{0.34}_{0.03}$&1.7&7.5&6.7\\
\hline
\end{tabular}
\begin{list}{Table Notes.}
\item $\ebv_O$ is the colour excess of the cluster's central region (Sect.~\ref{age}), while the bounds refer to
the range of reddening throughout the cluster field. Uncertainty in $\ebv_O$ is of the order of $\pm0.05$. 
Col.~14: \dgc\ calculated using $R_O=8.0$\,kpc (Reid \cite{Reid93}) as the distance of the Sun to the Galactic 
center. Col.~15: \dgc\ using $R_O=7.2$\,kpc (Bica et al. \cite{BBBO06}).
Uncertainties in \ds\ and \dgc\ are of the order of $\pm0.2$\,kpc.
\end{list}
\end{table*}

\begin{itemize}
\item One of the earliest studies of M\,52 is that of Pesch (\cite{Pesch60}) that using UBV photolectric 
photometry found a non-uniform reddening varying in the range $\ebv=0.51 - 0.81$, a  distance from 
the Sun $\ds=1.66$\,kpc and suggested an age similar to that of the Pleiades ($\sim135$\,Myr, according
to WEBDA).

\item Lindoff (\cite{Lind68}) derived an age of $\approx35$\,Myr for M\,52.

\item Schmidt (\cite{Schmidt77}) using uvby-$\beta$ photometry found $\ebv=0.45 - 0.53$.

\item Danford \& Thomas (\cite{DT81}) using uvby-$\beta$ photometry found $\ds=1.47$\,kpc,
a mean $\ebv=0.57$ and non-uniform reddening with \ebv\ in the northern half of the
cluster field in excess of 0.11 with respect to the southern half. They found an age 
intermediate between those of the $\alpha$ Persei cluster ($\sim70$\,Myr) and the 
Pleiades.

\begin{figure*}
\begin{minipage}[b]{0.50\linewidth}
\includegraphics[width=\textwidth]{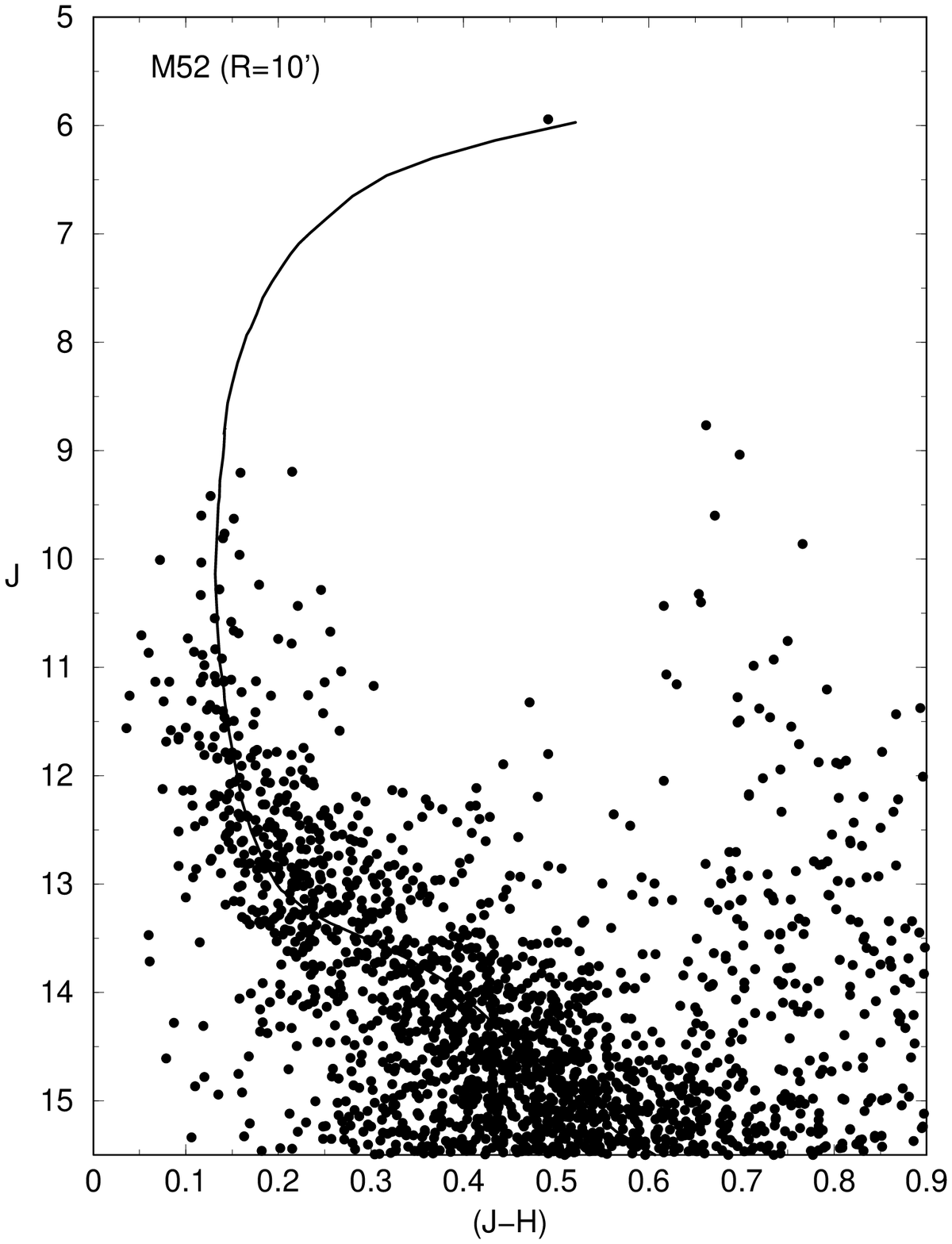}
\end{minipage}\hfill
\begin{minipage}[b]{0.50\linewidth}
\includegraphics[width=\textwidth]{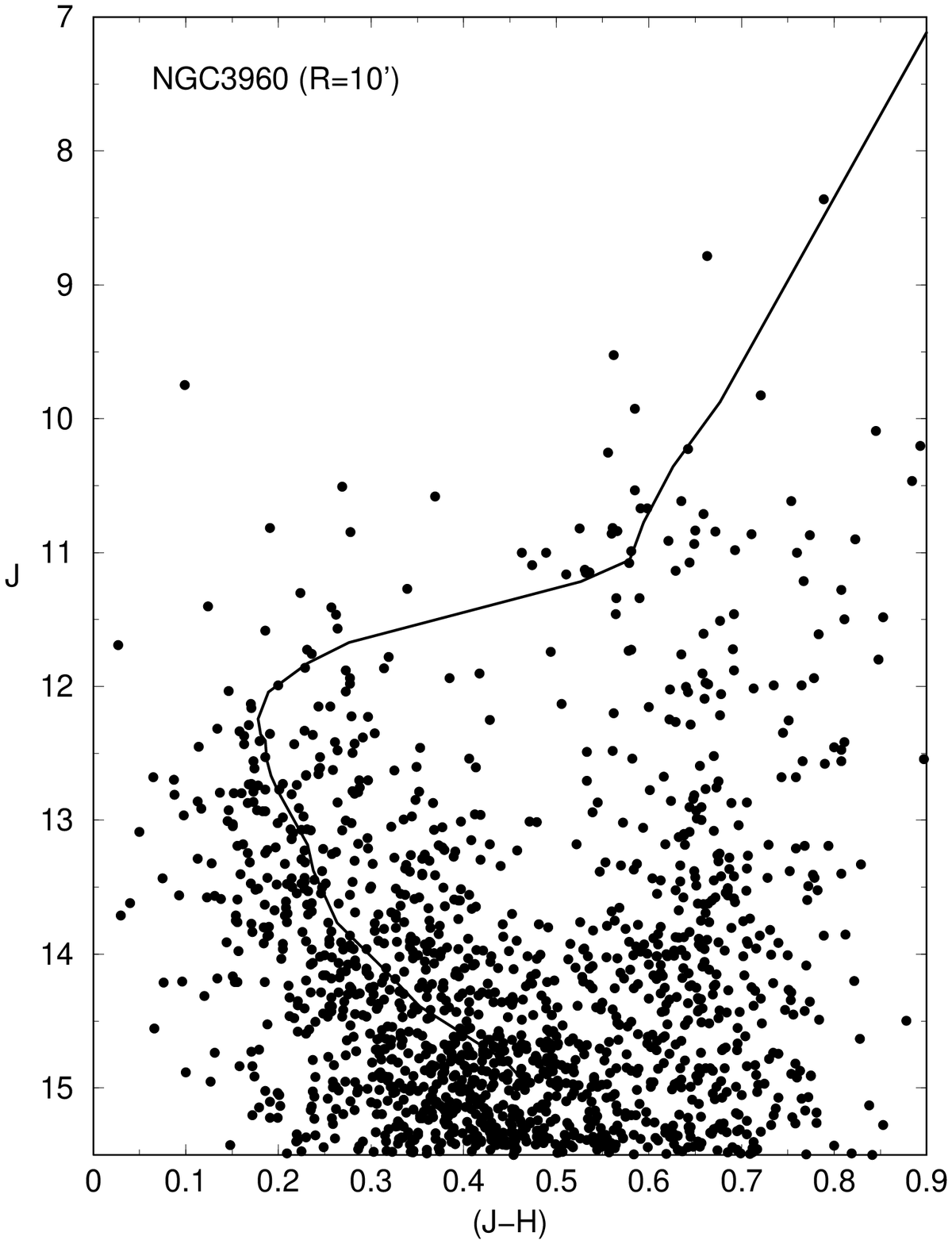}
\end{minipage}\hfill
\caption[]{$\jj\times\jh$ CMDs of the central 10\arcmin\ of M\,52 (left panel) and NGC\,3960 
(right panel). Solid lines: fiducial lines used to derive the differential reddening (Sect.~\ref{DiffRed}). 
Red stars belong mostly to the field (Sect.~\ref{FSD}). Error bars are shown in the right panels of 
Fig.~\ref{fig4}.}
\label{fig2}
\end{figure*}

\item Bruch \& Sanders (\cite{Bruch83}) applied an empirical calibration to convert relative
to absolute mass and estimated a cluster mass of $\approx518$\,\ms.

\item Lyng\aa\ (\cite{Lynga87}) classified M\,52 as Trumpler type II2r, and measured an angular 
diameter of $\rm D=12\arcmin$.
 
\item Kaltcheva (\cite{Kaltcheva90}) using uvby photometry found an age of $\approx96$\,Myr and
mean cluster reddening of $\ebv=0.57$.

\item Battinelli, Brandimarti \& Capuzzo-Dolcetta (\cite{Batt94}) using integrated UBV photometry
found an angular diameter of 12\arcmin, age $\approx35$\,Myr, $\ebv=0.57$, 202 member stars 
and mass $m\approx440$\,\ms.

\item Lotkin (\cite{Lotkin94}) estimated a main sequence (MS) turn-off age of $\rm\sim50\,Myr$ and 
$\ds=1.49$\,kpc.

\item Pandey et al. (\cite{Pandey2001}) using wide field ($40\arcmin\times40\arcmin$) CCD UBVIc 
photometry found variable reddening ($\ebv=0.46 - 0.80$), $\ds=1.38\pm0.07$\,kpc, and a spread 
in age from 30 to 100\,Myr. For stars with mass in the range 1.5---4.0\,\ms\ they found
MF $\left(\phi(m)=\frac{dN}{dm}\propto m^{-(1+\chi)}\right)$ slopes $\chi=1.07\pm0.08$ (inner region), 
$\chi=1.28\pm0.20$ (intermediate region), and $\chi=1.40\pm0.07$ (whole cluster). They concluded
that the presence of mass segregation is consistent with the cluster's age range.

\item Nilakshi et al. (\cite{Nilakshi2002}) using photometry from Palomar Observatory Sky Survey 
I plates found a Galactocentric distance $\dgc=9.2$\,kpc, core radius $\rc=1.59\pm0.08$\,pc 
and core stellar density $\rho_c=22.4\pm1.7\rm\,stars\,pc^{-2}$, halo radius $R_h=4.48$\,pc 
and halo stellar density $\rho_h=8.4\pm0.3\rm\,stars\,pc^{-2}$.

\item Kharchenko et al. (\cite{Kharchenko05}) derived an age of $\rm\approx69\,Myr$, 
$\ebv=0.65$, $\ds=1.42$\,kpc, $\rc=4.8\arcmin\ (\approx2\rm\,pc)$ and cluster radius of 
$10.2\arcmin\ (\approx4.2\rm\,pc)$. 
\end{itemize}

In WEBDA the central coordinates of M\,52 are (J2000) $\alpha=23^h24^m31^s$, and 
$\delta=+61^\circ35\arcmin29\arcsec$. However, the corresponding radial density 
profile built with stars having colours consistent with the cluster's (Sect.~\ref{struc}) presented 
a dip at $\rm R=0\arcmin$. Consequently, we searched for a new center by examining histograms for 
the number of stars in 0.5\arcmin\ bins of right ascension and declination. The coordinates that 
maximize the central density of stars are (J2000) $\alpha=23^h24^m42^s$, and 
$\delta=+61^\circ35\arcmin42\arcsec$. In what follows we refer to these optimized coordinates as 
the center of M\,52. Additional cluster parameters from WEBDA are given in Table~\ref{tab1}.

\subsection{NGC\,3960}
\label{N3960}

NGC\,3960 is a moderately-rich OC projected in the 4th quadrant (Table~\ref{tab1}). 
The $\rm20\arcmin\times20\arcmin$ DSS R image is in Fig.~\ref{fig1} (right panel).
Previous works on NGC\,3960 are listed below.

\begin{itemize}
\item Using BV and DDO photometry Janes (\cite{Janes81}) estimated the age of NGC\,3960 as similar
to that of the Hyades (0.5 -- 1\,Gyr), a mean reddening of $\ebv=0.29\pm0.02$ and a
metallicity $[Fe/H]=-0.30\pm0.06$.

\item Carraro, Ng \& Portinari (\cite{CNP98}) derived an age of 0.6\,Gyr and $[Fe/H]=-0.34$.

\item Prisinzano et al. (\cite{Prisinzano04}) with BVI photometry of a $34\arcmin\times33\arcmin$ field
centered on NGC\,3960 found an age in the range 0.9 -- 1.4\,Gyr, variable reddening $E(V-I)=0.21 - 
0.78$ ($\ebv=0.17 - 0.62$) with $E(V-I)=0.36$ ($\ebv=0.29$) at the cluster center, and 
an MF slope $\chi=1.81\pm0.84$.

\item Bragaglia et al. (\cite{Bragaglia06}) used UBVI CCD photometry to derive an age in the range
0.6 - 0.9\,Gyr, $\ds=2.1$\,kpc, $[Fe/H]=-0.12\pm0.05$, $\ebv=0.29\pm0.02$ with a differential
reddening $\Delta\ebv=\pm0.05$. They detected mass segregation and dynamical evaporation.

\item Similarly to M\,52 we had to recalculate the cluster's central coordinates with respect to those 
in WEBDA (Table~\ref{tab1}).
\end{itemize}

\begin{table*}
\caption[]{Differential reddening $\Delta\ebv$ in M\,52}
\label{tab2}
\renewcommand{\tabcolsep}{1.9mm}
\renewcommand{\arraystretch}{1.5}
\begin{tabular}{|c||c|c|c|c|c|c|c||c|}
\cline{1-8}
$\Delta Y (\arcmin)$&\multicolumn{7}{|c||}{$\Leftarrow\Delta X (\arcmin)\Rightarrow$}\\
\cline{2-9}
$\Downarrow$ &$+15.0$&$+10.0$&$+5.0$&$0.0$&$-5.0$&$-10.0$&$-15.0$&\avgE\\
\hline\hline
$+15.0$&$+0.22$&$+0.22$&$+0.06$&$+0.26$&$+0.29$&$+0.35$&$+0.13$&$0.22\pm0.10$\\
\hline
$+10.0$&$+0.11$&$+0.26$&$+0.24$&$+0.19$&$+0.16$&$+0.08$&$+0.11$&$0.16\pm0.07$\\
\hline
$+5.0$&$+0.05$&$+0.05$&$+0.11$&$+0.08$&$+0.18$&$+0.29$&$+0.21$&$0.14\pm0.09$\\
\hline
$~0.0$&$+0.08$&$-0.03$&$+0.05$&$~0.00$&$-0.02$&$+0.35$&$+0.06$&$0.07\pm0.13$\\
\hline
$-5.0$&$+0.08$&$+0.08$&$+0.06$&$-0.06$&$+0.06$&$+0.10$&$+0.03$&$0.05\pm0.05$\\
\hline
$-10.0$&$~0.00$&$-0.02$&$-0.06$&$+0.08$&$+0.06$&$+0.02$&$+0.11$&$0.04\pm0.08$\\
\hline
$-15.0$&$-0.03$&$-0.08$&$+0.03$&$+0.13$&$-0.05$&$0.06$&$-0.02$&$0.01\pm0.07$\\
\hline\hline
\avgE&$0.07\pm0.08$&$0.07\pm0.13$&$0.07\pm0.09$&$0.10\pm0.11$&$0.11\pm0.12$&$0.18\pm0.14$&$0.09\pm0.07$  \\
\cline{1-8}
\end{tabular}
\begin{list}{Table Notes.}
\item Uncertainties in $\Delta\ebv$ are of the order of $\pm0.05$. For absolute values: $\ebv=0.58+\Delta\ebv$. 
\end{list}
\end{table*}

\section{The 2MASS photometry}
\label{2mass}

VizieR\footnote{\em http://vizier.u-strasbg.fr/viz-bin/VizieR?-source=II/246} was used to 
extract \jj, \hh\ and \ks\ 2MASS photometry in a circular area with radius $\rm R=40\arcmin$ 
centered on the optimized coordinates of both clusters (Table~\ref{tab1}). As photometric quality 
constraints the extraction was restricted to stars {\em (i)} brighter than the 99.9\% Point 
Source Catalogue Completeness Limit\footnote{Following the Level\,1 Requirement, according to 
{\em\tiny http://www.ipac.caltech.edu/2mass/releases/allsky/doc/sec6\_5a1.html }}, $\jj=15.8$, 
$\hh=15.1$\ and $\ks=14.3$, respectively, and {\em (ii)} with errors in \jj, \hh\ and \ks\
smaller than 0.2\,mag. For reddening transformations we use the relations $A_J/A_V=0.276$, 
$A_H/A_V=0.176$, $A_{K_S}/A_V=0.118$, and $A_J=2.76\times\ejh$ (Dutra, Santiago \& Bica 
\cite{DSB2002}), assuming a constant total-to-selective absorption ratio $\rm R_V=3.1$. 

The $\jj\times\jh$ CMD of the central 10\arcmin\ of M\,52 is in Fig.~\ref{fig2} (left panel). Reflecting 
the relatively young age a prominent, nearly vertical MS is present, together with the bright star at
$\jj\approx6$ and $\jh\approx0.55$ (the red SG $\rm BD+60^{\circ}2532$). Because of 
its low latitude, field stars (mostly disc) contaminate the CMD particularly at faint magnitudes 
and red colours. 

The intermediate-age nature of NGC\,3960 is evident in the truncated MS and the presence of
a clump of red evolved stars at $\jh\approx0.6$ (right panel of Fig.~\ref{fig2}). However,
conspicuous field-star contamination precludes further speculations.

\subsection{Differential reddening}
\label{DiffRed}

\begin{figure*}
\begin{minipage}[b]{0.50\linewidth}
\includegraphics[width=\textwidth]{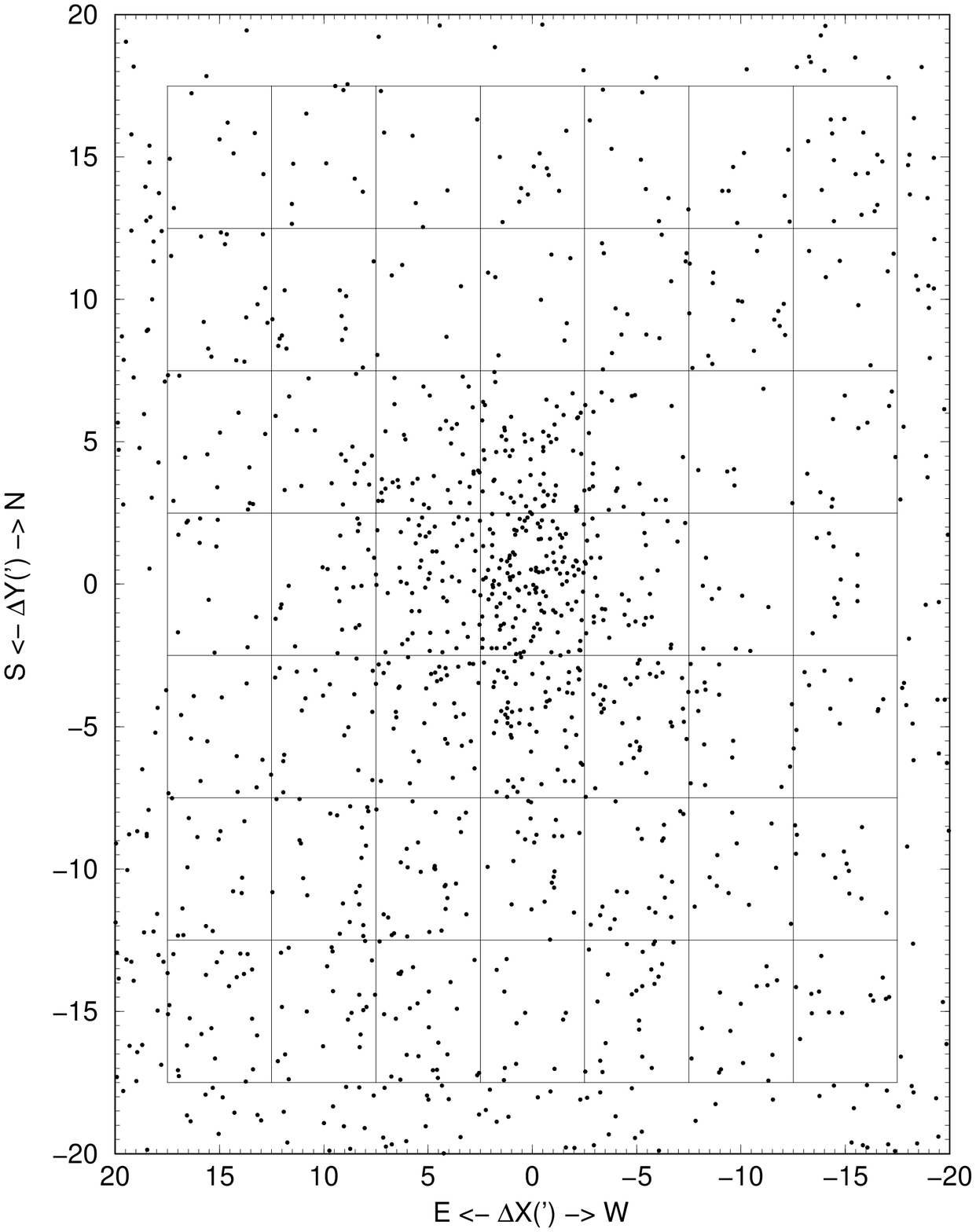}
\end{minipage}\hfill
\begin{minipage}[b]{0.50\linewidth}
\includegraphics[width=\textwidth]{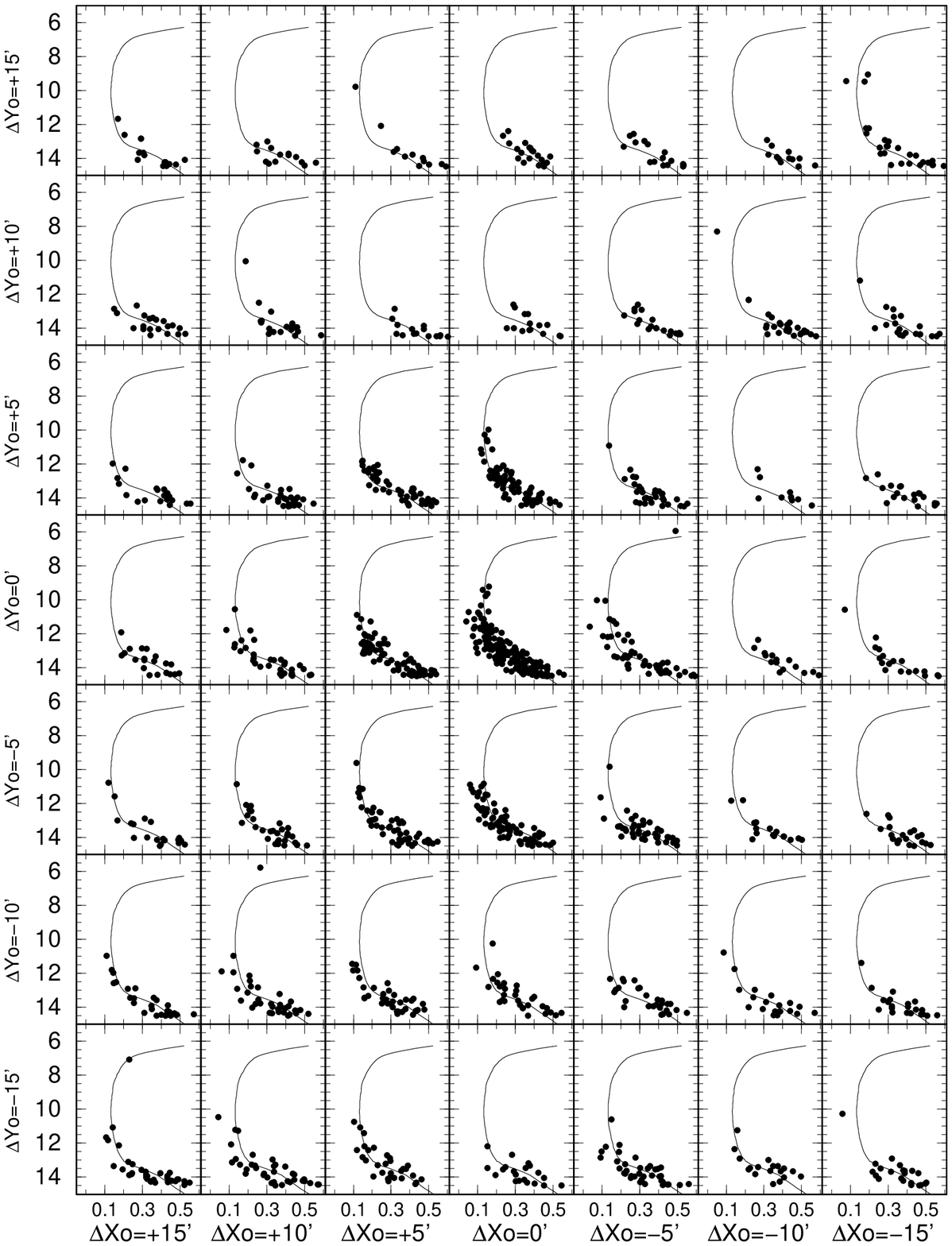}
\end{minipage}\hfill
\caption[]{Left panel: spatial grid used to derive the differential reddening in M\,52. Only stars 
with colours consistent with the cluster's (Sect.~\ref{struc}) are plotted here. North is up and East 
left. Right panel: $\jj\times\jh$ CMDs extracted from the spatial grid (left panel). The fiducial line 
of the central cell is shown in all panels as a continuous line. Spatial orientation follows that in 
the left panel. $\Delta X_O$ and $\Delta Y_O$ give the displacement along right ascension and declination
of the center of each cell with respect to the central cell.}
\label{fig3}
\end{figure*}

Previous work has shown that differential reddening is present in the fields of NGC\,3960 and
particularly M\,52 (Sect.~\ref{M52N3960}). Spatially variable reddening may introduce additional 
uncertainties in the analysis, particularly with respect to the intrinsic CMD morphology. To take 
this effect into account we divided the cluster fields into grids with cells of 5\arcmin\ width, 
along right ascension and declination, extending $17.5\arcmin$ in both directions. These 
dimensions correspond roughly to the cluster sizes (Sect.~\ref{struc}). The 5\arcmin\ cell dimension 
provides both spatial resolution and statistical significance in the number of stars in most cells. We 
illustrate this procedure in Fig.~\ref{fig3} for M\,52. For more accurate results we work only with 
the stars that remain after applying the colour-magnitude filter that discards stars with non-cluster 
colours (Sect.~\ref{struc}). In this plot the projected stellar density of the central region of M\,52 
clearly detaches from the outer regions. The corresponding $\jj\times\jh$ cell CMDs are shown in the 
right panel of Fig.~\ref{fig3}.

\begin{table*}
\caption[]{Differential reddening $(\Delta\ebv)$ in NGC\,3960}
\label{tab3}
\renewcommand{\tabcolsep}{1.9mm}
\renewcommand{\arraystretch}{1.5}
\begin{tabular}{|c||c|c|c|c|c|c|c||c|}
\cline{1-8}
$\Delta Y (\arcmin)$&\multicolumn{7}{|c||}{$\Leftarrow\Delta X (\arcmin)\Rightarrow$}\\
\cline{2-9}
$\Downarrow$ &$+15.0$&$+10.0$&$+5.0$&$0.0$&$-5.0$&$-10.0$&$-15.0$&\avgE\\
\hline\hline
$+15.0$&$-0.10$&$+0.06$&$+0.05$&$+0.16$&$-0.03$&$-0.02$&$+0.16$&$0.04\pm0.10$\\
\hline
$+10.0$&$+0.06$&$+0.11$&$~0.00$&$+0.03$&$+0.14$&$+0.16$&$+0.05$&$0.08\pm0.06$\\
\hline
$+5.0$ &$+0.03$&$~0.00$&$+0.03$&$+0.02$&$-0.06$&$+0.08$&$+0.21$&$0.04\pm0.08$\\
\hline
$~0.0$  &$+0.02$&$-0.02$&$-0.06$&$~0.00$&$-0.10$&$+0.08$&$+0.18$&$0.01\pm0.09$\\
\hline
$-5.0$ &$+0.14$&$+0.03$&$-0.03$&$+0.02$&$+0.05$&$+0.13$&$+0.05$&$0.06\pm0.06$\\
\hline
$-10.0$&$+0.16$&$+0.03$&$-0.06$&$+0.14$&$+0.08$&$+0.16$&$+0.03$&$0.08\pm0.08$\\
\hline
$-15.0$&$-0.08$&$+0.02$&$+0.21$&$+0.14$&$~0.00$&$+0.11$&$+0.18$&$0.08\pm0.10$\\
\hline\hline
\avgE&$0.03\pm0.10$&$0.03\pm0.04$&$0.02\pm0.09$&$0.07\pm0.07$&$0.01\pm0.08$&$0.10\pm0.06$&$0.12\pm0.08$ \\
\cline{1-8}
\end{tabular}
\begin{list}{Table Notes.}
\item Uncertainties in $\Delta\ebv$ are of the order of $\pm0.05$. For absolute values: $\ebv=0.13+\Delta\ebv$.
\end{list}
\end{table*}

Taking as reference the CMD of the central cell ($\Delta X_O=\Delta Y_O=0\arcmin$) we built a fiducial 
line for the MS and evolved stars that represents the mean \jh\ colour distribution in bins of 0.25 
mag in \jj. Because bright stars are scarce (particularly in M\,52), we used the respective 
Padova isochrone (Sect.~\ref{age}) as a guide. The CMD fiducial lines of the central region of both 
clusters are shown in Fig.~\ref{fig2}. For M\,52 it is superimposed on all cell CMDs (right panel of
Fig.~\ref{fig3}). To derive the differential reddening of a given cell with respect 
to the central one we applied a procedure that compares the observed CMD morphology with the fiducial line. 
Differences are assumed to be caused only by reddening. The procedure varies \ebv\ from $-0.8$ to $+0.8$ 
in steps of 0.02. The observed \jj\ and \hh\ magnitudes of all stars in the cell are corrected for each 
new value of \ebv, and the corresponding \jh\ colour is calculated. The chi-square is computed by summing 
the squared difference of \jh\ colour between observed point and fiducial line over all the stars in the 
cell. The excess reddening ($\Delta\ebv$) in each cell is the value of \ebv\ that produces the best match 
between fiducial line and observed colours. The resulting values of $\Delta\ebv$ for M\,52 are given in 
Table~\ref{tab2}, where we include as well the average values of $\Delta\ebv$ both for the right ascension 
and declination bins.

As previously indicated (e.g. Danford \& Thomas \cite{DT81}) there is a reddening 
gradient from the south towards the north through the field of M\,52. With
the present data we found that $\Delta\ebv$ ranges from $-0.08\pm0.05$ to $0.35\pm0.05$, and a 
least-squares fit to the data provides the relation $\Delta\ebv=(0.097\pm0.030)+(0.0067\pm0.00319)
\times\Delta Y$ from south to north, with a correlation coefficient $\rm CC=0.98$. With $\ebv=0.58\pm0.03$ 
derived for the central region of M\,52 (Sect.~\ref{age}), $\Delta\ebv$ values convert to absolute 
reddening values of $0.50\pm0.05$ and $0.93\pm0.05$, respectively. In right ascension, on the other hand, 
reddening is rather uniform, within uncertainties. The present values of the central and differential 
reddening agree with those of Danford \& Thomas (\cite{DT81}), Kaltcheva (\cite{Kaltcheva90}) and Pandey 
et al. (\cite{Pandey2001}).

A similar procedure was applied to NGC\,3960, and the respective values of $\Delta\ebv$ are given in
Table~\ref{tab3}. Contrarily to M\,52 the differential reddening is more uniform throughout 
the field of NGC\,3960, within the uncertainties. $\Delta\ebv$ ranges from $-0.10\pm0.05$ to
$0.21\pm0.05$ that, with the central value of $\ebv=0.13\pm0.03$ (Sect.~\ref{age}), convert 
to absolute reddening values of $0.03\pm0.05$ and $0.34\pm0.05$, respectively.

For the subsequent analyses we corrected the photometry for the differential reddening. We note, however,
that the correction was applied only to the stars located at $R\leq20\arcmin$ from the cluster center. 

\subsection{Field-star decontamination}
\label{FSD}

\begin{figure} 
\resizebox{\hsize}{!}{\includegraphics{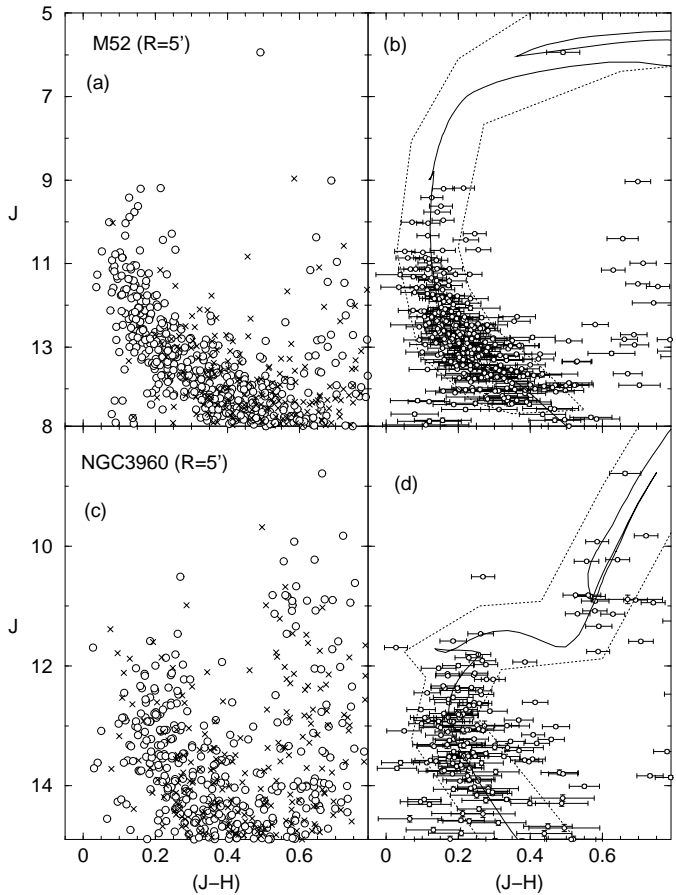}}
\caption[]{Panel (a): field stars ('x') are superimposed on the differential-reddening corrected 
CMD of the central 5\arcmin\ of M\,52. Extraction areas match. Panel (b): field-star decontaminated 
CMD. Solid line: 60\,Myr Padova isochrone with $\mMJ=11.3\pm0.1$ and $\ejh=0.18\pm0.01$. Bottom 
panels: same as the top panels for NGC\,3960. Solid line in panel (d): 1.1\,Gyr Padova isochrone with 
$\mMJ=11.3\pm0.1$ and $\ejh=0.04\pm0.01$. Colour-magnitude filters used to isolate MS/evolved stars 
are shown as dotted lines in panels (b) and (d).}
\label{fig4}
\end{figure}

Field stars are usually an important component of wide-field CMDs, particularly of low-latitude
star clusters and/or those projected against the bulge. Their presence in the fields of M\,52
and NGC\,3960 is clear (Fig.~\ref{fig2}), where they mimic an artificial sub-solar mass MS 
in M\,52 and a red sequence in NGC\,3960. To better evaluate this effect we show in the left
panels of Fig.~\ref{fig4} the $\jj\times\jh$ CMDs of the central 5\arcmin\ of M\,52 (top panel)
and NGC\,3960 (bottom). On both CMDs we superimpose the respective field-star contribution,
extracted from the region $39.686\leq R(\arcmin)\leq 40$ that corresponds to the same projected
area. Most of the faint stars ($\rm\jj\geq14.2$) in M\,52 and the red ones ($\jh\geq0.6$)
in both clusters are probably field stars.

\begin{table}
\caption[]{Cluster and field-star statistics}
\label{tab4}
\renewcommand{\tabcolsep}{2.3mm}
\begin{tabular}{lrrrrrrrr}
\hline\hline
&\multicolumn{3}{c}{M\,52}&&\multicolumn{3}{c}{NGC\,3960}\\
\cline{2-4}\cline{6-8}
$\Delta R$&$N_{obs}$&$f_{cl}$&$f_{fs}$& &$N_{obs}$&$f_{cl}$&$f_{fs}$ \\
(\arcmin)&(stars)&(\%)&(\%)&&(stars)&(\%)&(\%)\\
   (1)   &(2)    &(3) &(4) &&(5)    &(6) &(7)\\
\hline
0--3&313&63&37&&210&51&49\\
3--6&601&43&57&&434&28&72\\
6--9&726&21&79&&582&11&89\\
9--12&874&8&92&&803&9&91\\
12--15&1053&2&98&&927&0&100\\
Total field&3567&20&80&&2029&18&82\\
\hline
\end{tabular}
\begin{list}{Table Notes.}
\item Cols.~2 and 5: number of observed stars in the region. Cols.~3 and 6: fraction of
member stars. Cols.~4 and 7: fraction of field stars. Total field is $R=15\arcmin$ for M\,52 
and $R=12\arcmin$ for NGC\,3960.
\end{list}
\end{table}

To retrieve the intrinsic cluster-CMD morphology we use a field-star decontamination procedure
previously applied in the analysis of low-contrast (Bica \& Bonatto \cite{LowC05}), young 
embedded (Bonatto, Santos Jr. \& Bica \cite{BSJB05}), and young (Bonatto et al. \cite{BBOB06}) 
OCs. In the present case we apply this procedure to the differential-reddening corrected photometry.
As the offset field we take the region $\rm20\leq R(\arcmin)\leq40$, that is large enough to produce  
statistical representativity of the field-stars, both in magnitude and colours. The number of 
stars in both offset fields is $\approx15\,000$. The 
algorithm divides the CMD in colour/magnitude cells from which stars are randomly subtracted in a 
fraction consistent with the expected number of field stars. The remaining stars end up located in 
CMD regions where the stellar density presents a clear excess over the field and consequently they 
have a high probability of being cluster members. Because it 
actually excludes stars from the original files - thus artificially changing both the radial 
distribution of stars and LFs - we use field-star decontamination exclusively to derive the intrinsic 
CMD morphology. Further details are given in Bonatto et al. (\cite{BBOB06}). 

In Table~\ref{tab4} we summarise the results of the decontamination procedure that quantify the
presence of field stars throughout the fields of M\,52 and NGC\,3960. Member stars appear
to be found up to $\approx15\arcmin$ from the center of M\,52, while in NGC\,3960 they 
extend to $\approx12\arcmin$. The fraction of member stars in both fields is similar,
$\sim20\%$. The field-star decontamination was applied to the whole range
of colours and magnitudes (i.e. without discarding stars with non-cluster colours, Sect.~\ref{age}). 
Consequently, the above cluster dimensions are somewhat smaller than that suggested in Fig.~\ref{fig3} 
for M\,52 and those derived in Sect.~\ref{struc} for both OCs undergoing colour-magnitude filtered 
photometry. This effect is better illustrated with the observed and colour-magnitude filtered radial 
density profiles (Fig.~\ref{fig5}).

The field-star decontaminated CMDs of the region $R\leq5\arcmin$ are shown in the right panels 
of Fig.~\ref{fig4}, where error bars are included to show the magnitude of the photometric
uncertainties. As expected, most of the faint stars in M\,52 and red ones in NGC\,3960 were 
eliminated by the decontamination procedure. According to Fig.~\ref{fig4}, M\,52's 
CMD morphology is typical of a cluster younger than $\sim100$\,Myr with a well-developed MS spanning 
$\sim5$ magnitudes, while the old age of NGC\,3960 is reflected in the low turnoff and 
giant branch.

The colour spread in the MS of both clusters may be partly intrinsic and partly accounted for 
by residual reddening, since we could not compute differential reddening for the offset field 
($R\geq20\arcmin$). However, differential reddening effects in the near-IR are smaller
than in the optical, since $\ejh\approx0.3\times\ebv$. To further minimise this effect we work with 
colour-magnitude filters broad enough to include most of the colour spread of the high-probability 
member stars (right panels of Fig.~\ref{fig4}).

\subsection{Cluster age and distance from the Sun}
\label{age}
 
Cluster age is derived by means of solar-metallicity Padova isochrones (Girardi et al. 
\cite{Girardi2002}) computed with the 2MASS \jj, \hh\ and \ks\ filters\footnote{\em\tiny 
http://pleiadi.pd.astro.it/isoc\_photsys.01/isoc\_photsys.01.html. 2MASS transmission 
filters produced isochrones very similar to the Johnson ones, with differences of at most 
0.01 in \jh\ (Bonatto, Bica \& Girardi \cite{BBG2004}).}.

The red SG together with the rather well-defined low-MS of M\,52 reaching $\jj\approx14$ 
constrain the age to $60\pm10$\,Myr. Parameters derived from the isochrone fit are the observed 
distance modulus $\mMJ=11.3\pm0.1$ and colour excess $\ejh=0.18\pm0.01$, converting to $\ebv=0.58\pm0.03$. 
This age-solution is plotted in panel (b) of Fig.~\ref{fig4}. The absolute 
distance modulus is $\mMo=10.8\pm0.1$, resulting in $\rm\ds=1.4\pm0.2\,kpc$. MS stars are restricted 
to the mass range $1.3\leq m(\ms)\leq6.3$. The Galactocentric distance of M\,52 is $\dgc=8.7\pm0.2$\,kpc, 
using $R_O=8.0$\,kpc as the Sun's distance to the Galactic center (Reid \cite{Reid93}). However,
with the recently derived value of $R_O=7.2$\,kpc (based on updated parameters of globular clusters - 
Bica et al. \cite{BBBO06}), M\,52 is located $\dgc=7.9\pm0.2$\,kpc from the Galactic center. M\,52 is 
located $\approx0.7$\,kpc outside the Solar circle. 

For NGC\,3960 we derive an age of $1.1\pm0.1$\,Gyr, $\mMJ=11.3\pm0.1$ and $\ejh=0.04\pm0.01$, 
corresponding to $\ebv=0.13\pm0.03$, $\rm\ds=1.7\pm0.2\,kpc$ and $\dgc=7.5\pm0.2$\,kpc ($R_O=8.0$\,kpc) 
or $\dgc=6.7\pm0.2$\,kpc ($R_O=7.2$\,kpc). Either way, NGC\,3960 is $\approx0.5$\,kpc inside the Solar 
circle. Because of the 2MASS faint-magnitude limit (Sect.~\ref{2mass}), MS stars are detected in the range 
$0.95\leq m(\ms)\leq1.98$. This age solution is shown in panel (d) of Fig.~\ref{fig4}.

\subsection{Cluster structure}
\label{struc}

Cluster structure was inferred with the radial density profile (RDP), defined as the projected
number-density of MS/evolved stars around the center. The stars were selected by applying the 
respective colour-magnitude filters (right panels of Fig.~\ref{fig4}) to the differential-reddening 
corrected photometry (Sect.~\ref{DiffRed}). The use of colour-magnitude filters to discard 
foreground/background field objects was previously applied in the analysis of the OCs 
M\,67 (Bonatto \& Bica \cite{BB2003}), NGC\,3680 (Bonatto, Bica \& Pavani \cite{BBP2004}), 
NGC\,188 (Bonatto, Bica \& Santos Jr. \cite{BBS2005}), NGC\,6611 (Bonatto, Santos Jr. \& Bica 
\cite{BSJB05}), and NGC\,4755 (Bonatto et al. \cite{BBOB06}). To avoid oversampling near the center 
and undersampling for large radii, the RDPs were built by counting stars in concentric rings with 
radius $\Delta R=0.5\arcmin$ for $0\leq R(\arcmin)<5$, $\Delta R=1\arcmin$ for $5\leq 
R(\arcmin)<14$, $\Delta R=2\arcmin$ for $14\leq R(\arcmin)<30$ and $\Delta R=4\arcmin$ for 
$R\geq30\arcmin$. The residual background level corresponds to the average number of stars in the 
region $20\leq R(\arcmin)\leq 40$, resulting in $\sigma_{bg}=0.656\pm0.035\rm\,stars\,(\arcmin)^{-2}$
and $\sigma_{bg}=1.204\pm0.036\rm\,stars\,(\arcmin)^{-2}$, respectively for M\,52 and NGC\,3960.

\begin{figure} 
\resizebox{\hsize}{!}{\includegraphics{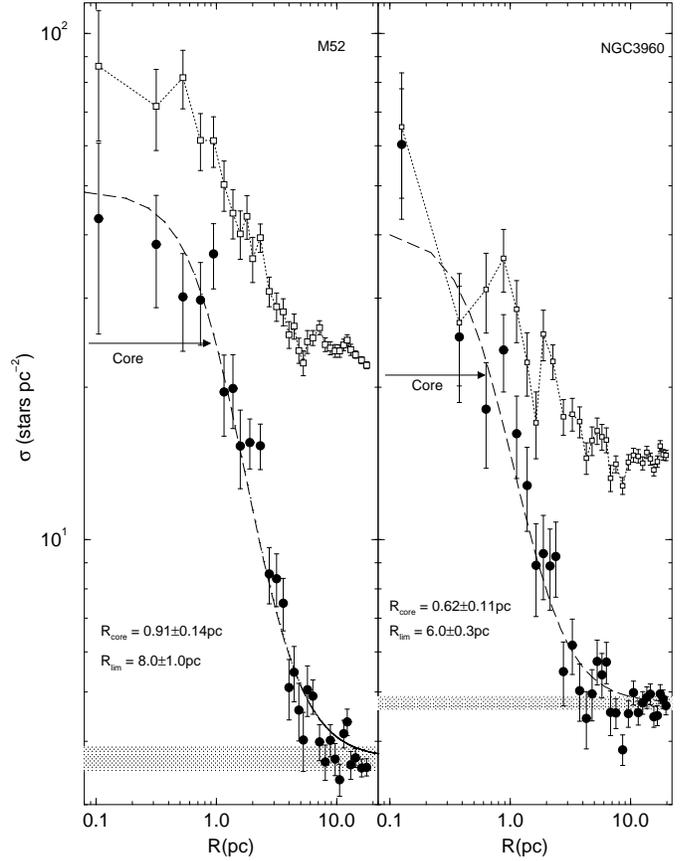}}
\caption[]{Filled circles: differential-reddening corrected $+$ colour-magnitude filtered radial 
density profiles (RDPs) of M\,52 (left panel) and NGC\,3960 (right 
panel). Dashed lines: two-parameter King profile where the core size is indicated. Shaded regions: 
stellar background level. For comparison purposes the RDPs prior to the colour-magnitude filtering
are shown as empty circles.}
\label{fig5}
\end{figure}

Fig.~\ref{fig5} shows the MS/evolved stars' RDP of both clusters. For absolute comparison between clusters 
the radius scale was converted to parsecs and the number-density of stars to $\rm stars\,pc^{-2}$\ using the 
distances derived in Sect.~\ref{age}. The statistical significance of the RDPs is indicated by the relatively
small $1\sigma$\ Poisson error bars. We also show in Fig.~\ref{fig5} the RDPs produced with photometry
prior to the colour-magnitude filtering. Clearly, the filtered profiles present less fluctuation and 
reveal more of the cluster structure than the observed ones. In particular, the observed profiles tend
to underestimate the cluster's extension.  

\begin{table}
\caption[]{Structural parameters.}
\begin{tiny}
\label{tab5}
\renewcommand{\tabcolsep}{1.25mm}
\renewcommand{\arraystretch}{1.4}
\begin{tabular}{lcccccc}
\hline\hline
Cluster&$\sigma_{bg}$&$\sigma_{0K}$&\rc&\rl&\rt \\
       &$\rm(stars\,pc^{-2})$&$\rm(stars\,pc^{-2})$&(pc)&(pc)&(pc)\\
(1)&(2)&(3)&(4)&(5)&(6)\\
\hline
&\multicolumn{5}{c}{Before differential-reddening correction}\\
\cline{2-6}
M\,52&$3.2\pm0.2$&$49\pm13$&$0.80\pm0.14$&$8.0\pm1.0$&---\\
NGC\,3960&$4.4\pm0.1$&$35\pm9$&$0.69\pm0.12$&$6.0\pm0.3$&--- \\
\cline{2-6}
&\multicolumn{5}{c}{After differential-reddening correction}\\
\cline{2-6}
M\,52&$3.7\pm0.2$&$45\pm10$&$0.91\pm0.14$&$8.0\pm1.0$&$13.1\pm2.2$\\
NGC\,3960&$4.8\pm0.2$&$36\pm9$&$0.62\pm0.11$&$6.0\pm0.8$&$10.7\pm3.7$ \\
\hline
\end{tabular}
\begin{list}{Table Notes.}
\item We express King's profile as $\sigma(R)=\sigma_{bg}+\sigma_{0K}/(1+(R/R_{\rm core})^2)$.
$\sigma_{0K}$ and \rc\ were allowed to vary while $\sigma_{bg}$ was kept fixed.
\end{list}
\end{tiny}
\end{table}

Structural parameters were derived by fitting the colour-magnitude filtered RDPs with the two-parameter 
King (\cite{King1966a}) surface density profile, which describes the intermediate and central regions of 
normal globular clusters (King \cite{King1966b}; Trager, King \& Djorgovski \cite{TKD95}). The fits 
were performed 
using a nonlinear least-squares fit routine that uses errors as weights. To minimise degrees of freedom
in the fit the background level ($\sigma_{bg}$) was kept constant, corresponding to the residual values
(see above). Parameters derived from the fit are King's central density of stars ($\sigma_{0K}$) and core 
radius (\rc). The resulting parameters are given in Table~\ref{tab5} and the best-fit solutions are superimposed 
on the colour-magnitude filtered RDPs (Fig.~\ref{fig5}). The limiting radius (\rl) of a cluster can be estimated 
by considering the fluctuations of the RDPs with respect to the background. \rl\ describes where the RDP merges 
into the background and for practical purposes most of the cluster stars are contained within $\rl$. M\,52
has core and limiting radii $\sim50\%$ and $\sim33\%$ larger than those of NGC\,3960, respectively 
(Table~\ref{tab5}).

We also give in Table~\ref{tab5} parameters derived from the RDPs built with differential reddening 
uncorrected photometry. The uncorrected values of $\sigma_{0K}$, \rc, and \rl\ agree with the corrected 
ones, within uncertainties.

The differential reddening corrected RDPs allowed us to compute the tidal radii of M\,52 and
NGC\,3960 using the three-parameter King (\cite{King1962}) model. The results are $\rt=13.1\pm2.2$\,pc 
($\sim1.6\times\rl$) and $\rt=10.7\pm3.7$\,pc ($\sim1.8\times\rl$). The present ratios $\rt/\rl$ agree 
with those derived for bright OCs in Bonatto \& Bica (\cite{BB2005}, that are in the range 1.4 --- 1.9.

The colour-magnitude filtering provided parameters of M\,52 to be compared with
those of Nilakshi et al. (\cite{Nilakshi2002}) and Kharchenko et al. (\cite{Kharchenko05}). The latter 
works do not apply such a filtering method. Probably because of the deep RDPs produced by the present 
method, our values of \rc\ and \rl\ for M\,52 are about 50\% smaller and larger, respectively, than those 
of Nilakshi et al. (\cite{Nilakshi2002}) and Kharchenko et al. (\cite{Kharchenko05}). 

King's profile provides an excellent analytical representation of the stellar RDP of M\,52 from the external
parts to the core. Since it follows from an isothermal (virialized) sphere, the close similarity of the stellar 
RDP with a King profile may suggest that the internal structure of M\,52 (particularly the core) has reached some 
level of energy equipartition after $\sim60$\,Myr (Sect.~\ref{age}). For NGC\,3960, on the other hand, King's 
profile fails to reproduce the innermost region, particularly in the core where there is a marked excess density 
of stars (Fig.~\ref{fig5}). In this region the RDP increases almost as a straight line towards the center. This 
fact may suggest post-core collapse in this $\sim1$\,Gyr OC (Sect.~\ref{age}), as seen in part of the globular 
clusters (Trager, King \& Djorgovski \cite{TKD95}). We  will return to this point in Sect.~\ref{CWODS}.

\section{Luminosity and mass functions}
\label{MF}

Luminosity and mass functions $\left(\phi(m)=\frac{dN}{dm}\right)$ for the core, halo and 
overall cluster are derived for M\,52 and NGC\,3960 following the methods presented in Bonatto
\& Bica (\cite{BB2005} and references therein). To maximize the statistical significance 
of field-star counts we take as the offset field the region $20\leq R(\arcmin)\leq40$, that lies 
$\geq5\arcmin$ and $\geq8\arcmin$ beyond the limiting radii of M\,52 and NGC\,3960
(Sect.~\ref{struc}). 

To isolate MS stars we apply the colour-magnitude filters (Fig.~\ref{fig4}) to the 
differential-reddening corrected photometries (Sect.~\ref{DiffRed}), resulting in the
MS ranges $8.8\leq\jj\leq14.2$ and $11.8\leq\jj\leq15.4$, respectively for M\,52 and 
NGC\,3960. The faint-magnitude limit of the MS in both clusters is brighter than that 
of the 99.9\% Completeness Limit (Sect.~\ref{2mass}). 

To take the residual field-star contamination into account we build LFs for each cluster region and 
offset field separately. \jj, H and \ks\ LFs are built by counting stars in magnitude bins from the 
respective faint magnitude limit to the turn-off, for cluster and offset field regions. Magnitude 
bins are wider in the upper MS than in the lower MS to avoid undersampling near the turn-off and 
oversampling at the faint limit. Corrections are made for different solid angles between offset 
field and cluster regions. Intrinsic LFs are obtained by subtracting the offset-field LFs 
(Fig.~\ref{fig7}); they are transformed into MFs using the mass-luminosity relations obtained from 
the respective Padova isochrones and observed distance modulii (Sect.~\ref{2mass}). These procedures 
are applied independently to the three 2MASS bands, and the final MFs combine the \jj, \hh\ and \ks\ 
MFs. MS mass ranges are $1.3\leq m(\ms)\leq6.1$ and $0.95\leq m(\ms)\leq1.98$, respectively for M\,52 
and NGC\,3960. We note that the 1.3\,\ms\ low-mass limit in M\,52 is consistent under contraction arguments
with an age of
$\approx60$\,Myr. Although expected to populate the MS of a $\approx1$\,Gyr-old OC, stars less massive 
than 0.95\,\ms\ cannot be detected in NGC\,3960 with 2MASS mostly because of the increasing fraction
of field-star contamination in that mass range (as already indicated by panel (d) of Fig.~\ref{fig4}).

\begin{figure} 
\resizebox{\hsize}{!}{\includegraphics{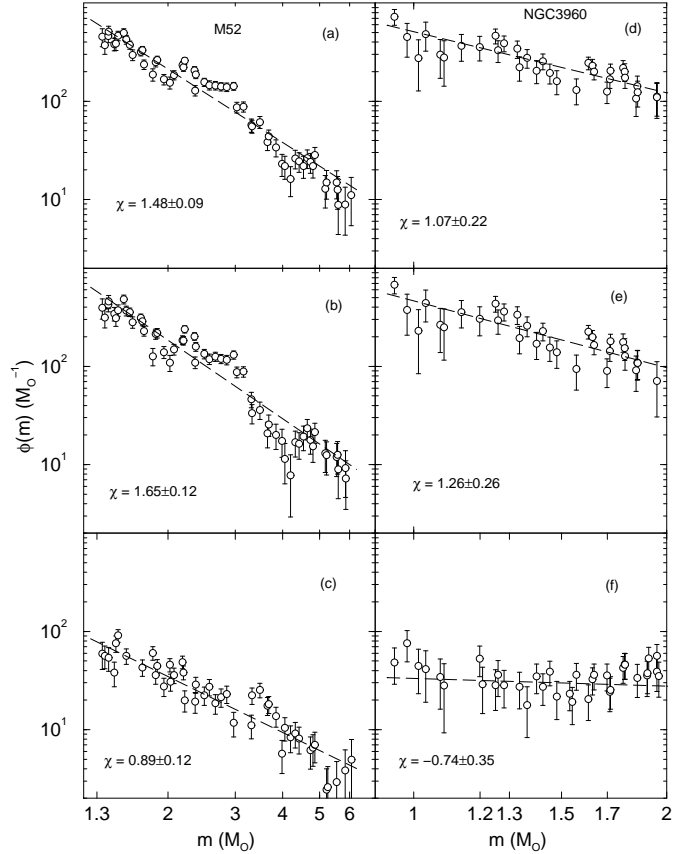}}
\caption[]{Overall (top panels), halo (middle) and core (bottom) mass functions of M\,52 (left panels)
and NGC\,3960 (right). Dashed lines: fits with $\phi(m)\propto m^{-(1+\chi)}$.}
\label{fig6}
\end{figure}

Fig.~\ref{fig6} shows the overall, halo and core MFs of both clusters together with the fits with the 
function $\phi(m)\propto m^{-(1+\chi)}$, and the MF slopes are given in Tables~\ref{tab6} and \ref{tab7}. 

Both the overall and halo MFs of M\,52 with slopes $\chi=1.48\pm0.09$ and $\chi=1.65\pm0.12$ are 
slightly steeper than Salpeter's (\cite{Salpeter55}) IMF ($\chi=1.35$). The core, however, is 
flatter with $\chi=0.89\pm0.12$. Within uncertainties the present  core, halo and overall MF 
slopes agree with those of Pandey et al. (\cite{Pandey2001}) for the equivalent mass range.

The overall MF of NGC\,3960 with slope $\chi=1.07\pm0.22$ is flatter than Salpeter's, while that of 
the halo ($\chi=1.26\pm0.26$) is similar. The core is very flat with $\chi=-0.74\pm0.35$. The present 
overall MF slope is significantly flatter than that of Prisinzano et al. (\cite{Prisinzano04}). In M\,52 
and particularly NGC\,3960, MF slopes increase from core to halo, which suggests mass segregation in 
the core. We will discuss this point further in Sect.~\ref{dyna}.
 
\subsection{Cluster mass}
\label{TM}

Parameters derived from the core, halo and overall MFs of M\,52 and NGC\,3960 are given in Tables~\ref{tab6} 
and \ref{tab7}. The number of evolved stars (col.~2) in each cluster region was obtained by integration of 
the respective field-star subtracted LF for stars earlier than the turnoff. Multiplying this number by the 
mass at the turn-off ($m\approx6$\,\ms\ in M\,52 and $m\approx2$\,\ms\ in NGC\,3960) yields an estimate of 
the mass stored in evolved stars (col.~3). The observed number of MS stars and corresponding mass (cols.~5 
and 6, respectively) were derived by integrating the MFs in the mass ranges $1.3\leq m(\ms)\leq6.1$ and 
$0.95\leq m(\ms)\leq1.98$, respectively for M\,52 and NGC\,3960. 

The total number (col.~7) and mass (col.~8) of the observed stars in M\,52 is computed by adding the corresponding 
values of the evolved and MS stars. We also give in Table~\ref{tab6} the projected mass density (col.~9) and 
volume mass density (col.~10). This total mass estimate for M\,52 should be
taken as a lower limit, since we are not computing the low-mass stars still in the process of reaching
the MS. On the other hand, upper-limits can be computed if we assume that stars less massive than 1.3\,\ms,
although not detected in the 2MASS photometry, are present in the cluster in the form of PMS stars. We do 
this by extrapolating the MFs down to the H-burning mass limit, $0.08\,\ms$ assuming  Kroupa's 
(\cite{Kroupa2001}) universal Initial Mass Function (IMF), in which $\chi=0.3\pm0.5$ for the range 
$0.08\leq m(\ms)\leq0.5$. For masses in the range $0.5\leq m(\ms)\leq1.3$ we use our values of $\chi$ which, 
within uncertainties are similar to the slope in Kroupa (\cite{Kroupa2001}) for the equivalent mass range. 
The extrapolated values of number, mass, projected and volume densities are given in cols.~7 to 10 of the third 
set of entries in Table~\ref{tab6}. According to this extrapolation the core may contain about twice the observed 
mass, while the total cluster mass may be a fraction of $\sim3$ larger than the observed value. 

Because of the 1.1\,Gyr age, we estimate the total mass locked up in stars in NGC\,3960 by taking into 
account all stars from the turnoff down to the H-burning mass limit, $0.08\,\ms$. For the core we 
extrapolate the flat MF down to $0.08\,\ms$. For the halo and overall cluster, on the other hand, we assume 
the universal Initial Mass Function (IMF) of Kroupa (\cite{Kroupa2001}). The results are given in Table~\ref{tab7}.

\begin{table*}
\caption[]{Parameters derived from the MFs of M\,52}
\label{tab6}
\renewcommand{\tabcolsep}{1.25mm}
\renewcommand{\arraystretch}{1.4}
\begin{tabular}{ccccccccccccccc}
\hline\hline
&\multicolumn{2}{c}{Evolved}&&\multicolumn{4}{c}{Observed MS}&&
\multicolumn{6}{c}{$\rm Evolved\ +\ Observed\ MS\ stars$}\\
\cline{2-3}\cline{5-8}\cline{10-15}
Region&$N^*$&$m$&&$\chi$&&$N^*$&\mobs&&$N^*$&$m$&$\sigma$&$\rho$&&$\tau$\\
&(stars)&(\ms)&&&&($10^2$stars)&($10^2\ms$)&& ($10^2$stars)&($10^2\ms$)&($\rm \ms\,pc^{-2}$)&($\rm \ms\,pc^{-3}$)\\
 (1)  & (2)   & (3)  &&(4)     && (5)    & (6)   && (7)    & (8)   &(9)     &(10)&&(11)\\
\hline
&\multicolumn{14}{c}{Before differential-reddening correction}\\
\cline{2-15}
Core& --- & --- &&$0.73\pm0.11$& &$0.9\pm0.1$&$2.2\pm0.4$
&&$0.9\pm0.1$&$2.2\pm0.4$&$109\pm17$&$102\pm16$&&$97\pm26$\\

Halo&$3\pm1$ &$18\pm6$ &&$1.69\pm0.11$& &$4.7\pm0.6$&$9.7\pm1.2$
 &&$4.7\pm0.6$&$9.8\pm1.3$&$4.9\pm0.6$&$0.46\pm0.06$&&---\\

Overall& $3\pm1$&$18\pm6$ &&$1.53\pm0.09$& &$5.3\pm0.6$&$11.3\pm1.3$ 
&&$5.4\pm0.7$&$11.5\pm1.3$&$5.7\pm0.6$&$0.5\pm0.1$&&$2.2\pm0.5$\\
\cline{2-15}
&\multicolumn{14}{c}{After differential-reddening correction}\\
\cline{2-15}

Core& --- & --- &&$0.89\pm0.12$& &$1.0\pm0.2$&$2.4\pm0.4$
&&$1.0\pm0.1$&$2.3\pm0.4$&$91\pm15$&$75\pm12$&&$78\pm20$\\

Halo&$3\pm1$ &$18\pm6$ &&$1.65\pm0.12$& &$4.6\pm0.6$&$10.0\pm1.4$
 &&$4.7\pm0.6$&$10.1\pm1.4$&$5.1\pm0.7$&$0.47\pm0.07$&&---\\

Overall& $3\pm1$&$18\pm6$ &&$1.48\pm0.09$& &$5.5\pm0.5$&$11.9\pm1.3$ 
&&$5.4\pm0.6$&$12.1\pm1.4$&$6.0\pm0.7$&$0.6\pm0.1$&&$2.2\pm0.6$\\ 
\cline{2-15}
&\multicolumn{2}{c}{}&&\multicolumn{4}{c}{}&&\multicolumn{6}{c}{$\rm Evolved\ +\ Observed\ +\ Extrapolated\ MS\ stars$}\\
\cline{2-3}\cline{5-8}\cline{10-15}

Core& --- & --- &&$0.89\pm0.12$& &$1.0\pm0.2$&$2.4\pm0.4$
&&$8.3\pm5.4$&$4.8\pm1.1$&$184\pm41$&$152\pm34$&&$13\pm8$\\

Halo&$3\pm1$ &$18\pm6$ &&$1.65\pm0.12$& &$4.6\pm0.6$&$10.0\pm1.4$
 &&$82\pm60$&$34\pm11$&$17\pm6$&$1.6\pm0.5$&&---\\

Overall& $3\pm1$&$18\pm6$ &&$1.48\pm0.09$& &$5.5\pm0.5$&$11.9\pm1.3$ 
&&$88\pm64$&$38\pm12$&$19\pm6$&$1.8\pm0.6$&&$0.2\pm0.1$\\

\hline\hline
\end{tabular}
\begin{list}{Table Notes.}
\item MS mass range: $1.3-6.1$\,\ms. Col.~4: MF slope. Col.~11: dynamical-evolution parameter 
$\tau={\rm age}/t_{\rm rel}$. 
\end{list}
\end{table*}

We also give in Tables~\ref{tab6} and \ref{tab7} parameters computed without the
differential-reddening correction. The results are not very sensitive to this correction, at least
in the near-IR.

The present mass determinations for M\,52 are a factor $\approx2$ (observed) and
$\approx6$ (extrapolated) larger than that of Bruch \& Sanders (\cite{Bruch83}). The difference
in the mass estimates can be accounted for by different analytical methods. In the present case
we have based our results on colour-magnitude filtered photometry, while Bruch \& Sanders (\cite{Bruch83}) 
estimated mass according to an empirical relation that converts relative to absolute masses.

\begin{table*}
\caption[]{Parameters derived from the MFs of NGC\,3960}
\label{tab7}
\renewcommand{\tabcolsep}{1.05mm}
\renewcommand{\arraystretch}{1.4}
\begin{tabular}{ccccccccccccccc}
\hline\hline
&\multicolumn{2}{c}{Evolved}&&\multicolumn{4}{c}{Observed MS}&&
\multicolumn{6}{c}{$\rm Evolved\ +\ Observed\  +\ Extrapolated\ MS\ stars$}\\
\cline{2-3}\cline{5-8}\cline{10-15}
Region&$N^*$&$m$&&$\chi$&&$N^*$&\mobs&&$N^*$&$m$&$\sigma$&$\rho$&&$\tau$\\
&(stars)&(\ms)&&&&($10^2$stars)&($10^2\ms$)&& ($10^2$stars)&($10^2\ms$)&($\rm \ms\,pc^{-2}$)&($\rm \ms\,pc^{-3}$)\\
 (1)  & (2)   & (3)  &&(4)     && (5)    & (6)   && (7)    & (8)   &(9)     &(10)&&(11)\\
\hline
&\multicolumn{14}{c}{Before differential-reddening correction}\\
\cline{2-15}
Core&$5\pm1$&$10\pm2$&&$-0.75\pm0.31$&& $0.3\pm0.1$&$0.5\pm0.1$&&$0.8\pm0.1$&$0.8\pm0.1$
&$48\pm6$&$52\pm6$&&$2300\pm520$\\

Halo&$22\pm5$&$44\pm10$&&$1.40\pm0.26$&& $2.5\pm0.3$&$3.3\pm0.4$&&$41\pm31$&$14.4\pm5.7$
&$13\pm5$&$1.6\pm0.6$&&---\\

Overall&$25\pm6$&$50\pm12$&&$1.16\pm0.23$&& $2.8\pm0.3$&$3.7\pm0.4$&&$38\pm27$&$14.1\pm5.1$
&$12\pm4$&$1.6\pm0.6$&&$10\pm6$ \\
\cline{2-15}
&\multicolumn{14}{c}{After differential-reddening correction}\\
\cline{2-15}

Core&$5\pm1$&$10\pm2$&&$-0.74\pm0.35$&& $0.3\pm0.1$&$0.5\pm0.1$&&$2.5\pm1.7$&$1.1\pm0.3$
&$94\pm28$&$114\pm33$&&$1010\pm604$\\

Halo&$22\pm5$&$44\pm10$&&$1.26\pm0.26$&& $2.3\pm0.3$&$3.1\pm0.4$&&$35\pm25$&$12.6\pm4.7$
&$11.2\pm4.2$&$1.4\pm0.5$&&---\\

Overall&$25\pm6$&$50\pm12$&&$1.07\pm0.22$&& $2.7\pm0.3$&$3.6\pm0.4$&&$34\pm24$&$13.0\pm4.5$
&$11.5\pm4.0$&$1.4\pm0.5$&&$11\pm7$ \\

\hline\hline
\end{tabular}
\begin{list}{Table Notes.}
\item Same as Table~\ref{tab6} for NGC\,3960. Detected MS mass range: $0.95-1.98$\,\ms. 
\end{list}
\end{table*}

\section{Dynamical state}
\label{dyna}

\subsection{M\,52}
\label{dynam52}

Within uncertainties, the overall MF slope ($\chi=1.48\pm0.09$) in the mass range $1.3\leq 
m(\ms)\leq6.1$ agrees with that of a Salpeter ($\chi=1.35$) IMF. However, the MF slopes present 
significant spatial variations throughout the cluster, being flat ($\chi=0.89\pm0.12$) in the core 
and rather steep ($\chi=1.65\pm0.12$) in the halo (Table~\ref{tab6} and Fig.~\ref{fig6}). In older 
clusters such spatial variation of MF slope can be accounted for by dynamical mass segregation, in 
the sense that low-mass stars originally in the core are transferred to the cluster's outskirts, while 
massive stars sink in the core. This process produces a flat core MF and a steep one in the halo (e.g. 
Bonatto \& Bica \cite{BB2005} and references therein).

Mass segregation in a star cluster scales with the relaxation time $\tr=\frac{N}{8\ln N}\tcr$, 
where $\tcr=R/\sigma_v$ is the crossing time, N is the (total) number of stars and $\sigma_v$\ 
is the velocity dispersion (Binney \& Tremaine \cite{BinTre1987}).  The relaxation time is the 
characteristic time-scale for a cluster to reach some level of energy equipartition. For a 
typical $\sigma_v\approx3\,\kms$ (Binney \& Merrifield \cite{Binney1998}) we obtain $\tr(\rm 
overall)=27\pm4$\,Myr and $\tr(\rm core)=0.8\pm0.2$\,Myr for the observed values; using the 
extrapolated values we obtain $\tr(\rm overall)=311\pm204$\,Myr and $\tr(\rm core)=4.5\pm2.6$\,Myr. 
The $\approx60$\,Myr age of M\,52 (Sect.~\ref{2mass}) corresponds to $\sim78\times\tr(\rm core)$ 
(observed) or $\sim13\times\tr(\rm core)$ (extrapolated). Thus, some degree of MF slope flattening 
in the core is to be expected. However, the ratio cluster age to \tr\ drops to $\sim2.2$ (observed) 
or $\sim0.2$ (extrapolated) for the overall cluster, consistent with the Salpeter slope and absence 
of large-scale mass segregation. 

\subsection{NGC\,3960}
\label{dyna3960}

The flat core MF of NGC\,3960 ($\chi=-0.74\pm0.35$) seems consistent with the relaxation time $\tr(\rm
core)=1.1\pm0.7$\,Myr, which corresponds to $\sim10^{-3}$ of the cluster age. For the overall cluster 
we obtain $\tr(\rm overall)=100\pm63$\,Myr, corresponding to $\sim0.1$ of the cluster age,  which might 
explain its sub-Salpeter ($\chi=1.07\pm0.22$) MF. Large-scale mass segregation through the body of 
NGC\,3960 is expected to have occurred, as well as low-mass star evaporation. NGC\,3960 is dynamically 
more advanced than M\,52 (Sect.~\ref{CWODS}). 

\section{Diagnostic diagrams of open cluster parameters}
\label{CWODS}

Bonatto \& Bica (\cite{BB2005}) derived a set of parameters related to the structure and
dynamical evolution of OCs in different dynamical states. We analysed
nearby OCs with ages in the range $70-7\,000$\,Myr and masses in $400-5\,300$\,\ms, 
following most of the present methodology. The evolutionary parameter $\tau=\rm{age}/\tr$ 
(col.~11 of Tables~\ref{tab6} and \ref{tab7}) was found to be a good indicator of dynamical states. 
In particular, significant flattening in core and overall MFs due to dynamical effects such as mass 
segregation is expected to occur for $\tau_{\rm core}\geq100$ and $\tau_{\rm overall}\geq7$, respectively. 
To these clusters were added the very young OC (age $\sim1.3$\,Myr) NGC\,6611 (Bonatto, Santos 
Jr. \& Bica \cite{BSJB05}) and the young (age $\sim14$\,Myr) NGC\,4755 (Bonatto et al. \cite{BBOB06}).

\begin{figure} 
\resizebox{\hsize}{!}{\includegraphics{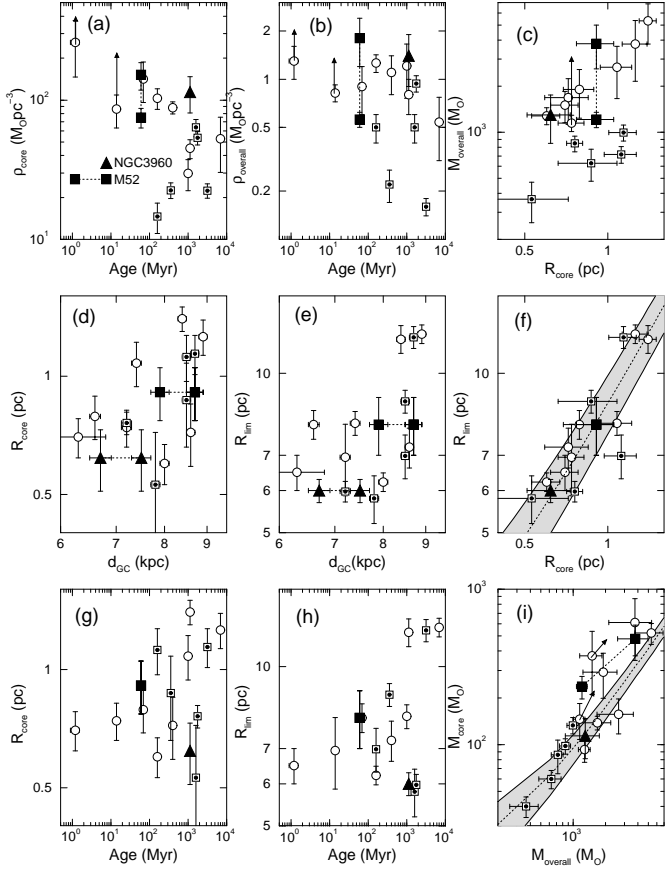}}
\caption[]{Relations involving structural parameters of OCs. Open circles: clusters 
more massive than 1\,000\,\ms. Squares with a dot: $m<1\,000$\,\ms. Dotted lines: least-squares 
fits to nearby clusters. Square: M\,52. Triangle: NGC\,3960. Shaded areas: $1\sigma$ borders 
of least-squares fits. Arrows indicate lower-limit estimates of mass, density and evolutionary 
parameter. Observed and extrapolated values of mass, density and evolutionary parameter of M\,52 
are shown. In panels (d) and (e) we show both determinations of \dgc.}
\label{fig7}
\end{figure}

\subsection{Structural parameters}
\label{StrucPar}

Structural parameters are dealt with in Fig.~\ref{fig7}. Because the observed values of mass, density 
and evolutionary parameters of M\,52 are lower limits, we plot as well the corresponding extrapolated
values.  For NGC\,3960 we only consider the extrapolated values. The core density of NGC\,3960, 
and particularly the extrapolated one of M\,52, follow the trend presented by massive OCs for the 
corresponding ages (panel a). A similar trend occurs for the overall 
density (panel b), particularly for the extrapolated value for M\,52. Both clusters follow the 
massive OC relation involving overall mass and core radius (panel c). Panels (d) and (e) appear
to suggest a correlation of \rc\ and \rl\ with \dgc, but the scatter of points precludes conclusions.
M\,52 and NGC\,3960 fit in the observed correlation of core and limiting radii (panel f). About $2/3$ 
of the sample OCs - including M\,52 - seem to define a trend of increasing core radius with age (panel g), 
while the remaining ones - including NGC\,3960 - appear to follow a sequence of decreasing \rc\ with 
age. Similar trends occur in the relation \rl\ with age (panel h). NGC\,3960 fits in the 
relation of core and overall mass (panel i), while M\,52 agrees at the $1\sigma$ level. Further details 
on parameter correlations are given in Bonatto \& Bica (\cite{BB2005}).

\begin{figure} 
\resizebox{\hsize}{!}{\includegraphics{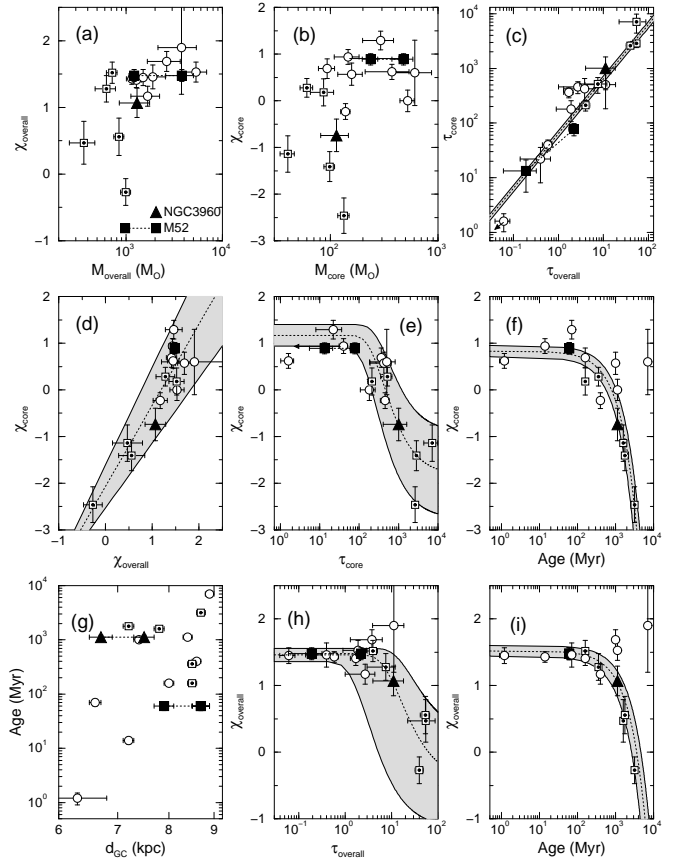}}
\caption[]{Relations involving evolutionary parameters of open clusters. Symbols as in 
Fig.~\ref{fig7}. Panels (f) and (i): fit of the linear-decay function $\chi(t)=\chi_o-t/t_f$.}
\label{fig8}
\end{figure}

\subsection{Dynamical parameters}
\label{DynaPar}

Dynamical-evolution parameters of M\,52 and NGC\,3960 are compared with those of other OCs in 
different dynamical states in Fig.~\ref{fig8}. The Salpeter-like overall MF slope of M\,52 is consistent 
with the relation of $\chi_{\rm overall}\times M_{\rm overall}$ for massive clusters (panel a); the same 
is valid for NGC\,3960 at the $1\sigma$ level. The core MF slope of M\,52 is consistent with the expected 
one for its core mass (panel b) while the core of NGC\,3960 seems  significantly flatter.
Both clusters fit in the relations involving $\tau_{\rm overall}\times\tau_{\rm core}$ (panel c), 
$\chi_{\rm overall}\times\chi_{\rm core}$ (panel d),
$\tau_{\rm core}\times\chi_{\rm core}$ (panel e),  and $\tau_{\rm overall}\times \chi_{\rm overall}$ 
(panel h). The distribution of Galactocentric distances in terms of age (panel g) indicates a correlation, 
as suggested by Lyng\aa\ (\cite{Lynga82}).  This correlation probably reflects the average tidal
disruption effect on clusters, which increases for smaller Galactocentric distances (e.g. Boutloukos \& 
Lamers \cite{BL2003}; Lamers et al. \cite{Lamers2005}).  

Panels (f) and (i) suggest a systematic flattening with age of the core and overall MFs, particularly 
for the less-massive OCs. The deviant object is the $\approx7$\,Gyr, massive OC NGC\,188 whose orbit 
avoids the inner regions of the Galaxy for most of the time (Bonatto, Bica \& Santos Jr. \cite{BBS2005} 
and references therein). The core and overall MFs of M\,52 have slopes consistent with those of other 
massive OCs of similar age. NGC\,3960, on the other hand, has core and overall MFs significantly flatter 
than those expected for its mass, although they may be consistent with its age (see below). Except for 
NGC\,188 the age dependence of the core and overall MF slopes can be parameterised by the linear-decay 
function $\chi(t)=\chi_o-t/t_f$, where $\chi_o$ represents the MF slope in the early phases and $t_f$ 
is the flattening time scale. For the core MF we derive $\chi_o=0.83\pm0.12$ and $t_f=0.9\pm0.1$\,Gyr 
(correlation coefficient $\rm CC=0.87$); the overall values are $\chi_o=1.52\pm0.08$ and $t_f=2.4\pm0.5$\,Gyr 
($\rm CC=0.79$). These relations suggest that OCs in general are formed with flat core and Salpeter-like 
overall MFs. At cluster birth the core MF already is much flatter than the
overall one, similarly to what was observed in e.g. NGC\,6611 (Bonatto, Santos Jr. \& Bica \cite{BSJB05}). 
Such early core flattening may be partly linked to primordial processes related to molecular-cloud fragmentation. 
Because of the cumulative effects of mass segregation, additional core MF flattening becomes significant after 
$\sim100$\,Myr and proceeds on an approximately linear time scale of $\sim0.9$\,Gyr (panel f). Flattening 
in the overall MF takes $\sim3\times$ longer to become evident and proceeds at a slower pace (panel i) than in the 
core, which probably reflects the average time scale associated with low-mass evaporation, tidal stripping 
and encounters with giant molecular clouds. Compared to the overall cluster, harder physical conditions 
in the core such as high density and small relaxation time account for the fast evolution rate (Bonatto 
\& Bica \cite{BB2005} and references therein).

The present MF flattening sequences - particularly the overall one - are defined mostly by the less-massive 
clusters, including NGC\,3960. At least for $\rm {ages}\leq3.2$\,Gyr, a linear-decay function 
provides a good analytical representation of the flattening, both for the core and overall MFs. It remains 
to determine the behaviour of the flattening in older clusters, whether it keeps linear until dissolution
into the field (particularly for the less-massive OCs) or saturates at some threshold value. A larger sample 
especially containing old OCs of any mass is required to {\em (i)} better constrain the analytical form of 
the less-massive OC flattening function, {\em (ii)} check whether the core and overall MFs of the massive OCs 
flatten in a similar way as those of the less-massive ones, and {\em (iii)} compute the core and overall MF 
flattening time scales.
 
\section{Concluding remarks}
\label{Conclu}

In this paper we analyzed the structure and distribution of stars in the OCs M\,52 and NGC\,3960 
with \jj, \hh\ and \ks\ 2MASS photometry, restricted to stars with observational uncertainties 
$\epsilon_{J,H,K_S}<0.2$\,mag. To derive cluster parameters we took into account the differential 
reddening and applied field-star decontamination and colour-magnitude filters. 

Compared to previous works we found differences in some parameters, such as cluster radius,
mass and MF slopes. Most of those studies were based on spatial and magnitude-limited data. However, 
the discrepancies can be largely accounted for by the absence of a proper field-star 
decontamination. 

Reddening throughout the field of M\,52 varies from $\ebv\approx0.50$ to $\ebv\approx0.93$, with a marked
gradient from south to north and a near uniformity in right ascension. For NGC\,3960 we derive
the range $\ebv\approx0.03 - 0.34$, uniform both in right ascension and declination. The projected 
area considered in both fields was $35\arcmin\times35\arcmin$.

For M\,52 we derived an age of $60\pm10$\,Myr, absolute distance modulus $\mMo=10.8\pm0.1$, and
distance from the Sun $\ds=1.4\pm0.2$\,kpc. For NGC\,3960 we found an age of $1.1\pm0.1$\,Gyr,
$\mMo=11.2\pm0.1$, and $\ds=1.7\pm0.2$\,kpc. M\,52 is located $\approx0.7$\,kpc outside the Solar 
circle, while NGC\,3960 is $\approx0.5$\,kpc inside it.

Radial density profiles built with colour-magnitude filtered photometry are much deeper and yield
more constrained OC structural parameters than the observed ones. In particular, King's model 
provided an excellent fit to the differential-reddening corrected radial density profile of M\,52, from
the external parts to the inner region. We found core and limiting radii of $\rc=0.91\pm0.14$\,pc and
$\rl=8.0\pm1.0$\,pc. The inner region of NGC\,3960, on the other hand, presents a clear excess of the 
stellar density over King's profile, which may suggest post-core collapse in this 1.1\,Gyr OC. Parameters 
derived are $\rc=0.62\pm0.11$\,pc and $\rl=6.0\pm0.8$\,pc. The tidal radii of M\,52 and NGC\,3960,
computed with the differential-reddening corrected RDPs are $\rt=13.1\pm2.2$\,pc and $\rt=10.7\pm3.7$\,pc.

The present use of colour-magnitude filtered photometry to build RDPs excludes a large fraction
of non-cluster stars. As a result, observed and colour-magnitude filtered RDPs present significant 
differences (especially depth and shape) as shown in Fig.~\ref{fig5}. Because of the small residual 
field-star contamination in the present RDPs, they can be used to derive more realistic intrinsic 
structural parameters. Nilakshi et al. (\cite{Nilakshi2002}) built RDPs using a photographic database 
(DSS photometry). They pointed out that the angular size was difficult to determine accurately due to 
the weak contrast between cluster and field stars. Kharchenko et al. (\cite{Kharchenko05}) derived 
parameters based on stars with available proper motion, which restricts the analysis to brighter stars. 
For evolved clusters in which a significant fraction of low-mass stars has already been displaced from 
the center outwards, their brighter cutoff photometry may explain differences, together with lower 
statistics for a smaller sample of  cluster stars. 

The mass stored in observed stars ($1.3\leq m(\ms)\leq6.1$) in M\,52 is $m_{\rm overall}\approx1.2\times10^3$\,\ms;
the core contains $m_{\rm core}\approx240$\,\ms. Extrapolating to the H-burning limit with Kroupa's IMF these values
become $m_{\rm overall}\approx3.8\times10^3$\,\ms\ and $m_{\rm core}\approx480$\,\ms. For NGC\,3960 we derived
$m_{\rm overall}\approx1.3\times10^3$\,\ms\ and $m_{\rm core}\approx114$\,\ms.

We conclude that both structurally and dynamically, the core and overall parameters of M\,52 
are consistent with those expected of an OC more massive than 1\,000\,\ms\ and $\approx60$\,Myr 
old. The somewhat flat core MF slope ($\chi_{\rm core}\approx0.89$) as compared to the steeper 
one at the halo ($\chi_{\rm halo}\approx1.65$) and the Salpeter-like overall MF 
($\chi_{\rm overall}\approx1.48$) suggests mild mass segregation in the inner region. This 
picture is consistent with the low values of the core/overall dynamical-evolution parameters, 
$\tau_{\rm core}\approx13 - 78$ and $\tau_{\rm overall}\approx0.2 - 2.2$. 

The parameters of NGC\,3960, on the other hand, suggest that this OC - particularly the core - has 
reached an advanced dynamical state characterised by $\tau_{\rm core}\sim10^3$ and 
$\tau_{\rm overall}\sim11$, respectively for the core and overall cluster. NGC\,3960 presents evident 
signs of mass segregation in the core/halo 
region, with a very flat MF slope ($\chi_{\rm core}\approx-0.74$) and a steeper one at the halo 
($\chi_{\rm halo}\approx1.26$). Besides, the somewhat flat (compared to Salpeter) overall MF 
($\chi_{\rm overall}\approx1.07$) suggests low-mass star evaporation through the cluster rim. 
Tidal disruption due to the Galactic field may have boosted the dynamical evolution of NGC\,3960, 
since it is located $\approx0.5$\,kpc inside the Solar circle.

Changes in parameters of M\,52 and NGC\,3960 because of differential-reddening correction are 
not significant, at least in the near-IR and/or because of the reddening ranges $\ebv=0.50 - 0.93$ 
(M\,52) and $\ebv=0.03 - 0.34$ (NGC\,3960). With respect to the analytical methods we conclude that 
colour-magnitude filters are essential to improve cluster parameters, particularly those related to 
structure. 

\begin{acknowledgements}
{We thank the referee, Dr. Nate Bastian, for helpful suggestions.}
This publication makes use of data products from the Two Micron All Sky Survey, which 
is a joint project of the University of Massachusetts and the Infrared Processing and 
Analysis Center/California Institute of Technology, funded by the National Aeronautics 
and Space Administration and the National Science Foundation. This research has made use 
of the WEBDA database, operated at the Institute for Astronomy of the University of Vienna.
CB and EB aacknowledge support from the Brazilian Institution CNPq.
\end{acknowledgements}

%

\end{document}